\documentclass[a4paper,fleqn,usenatbib]{mnras}

\usepackage[T1]{fontenc}
\usepackage{ae,aecompl}
 \usepackage{relsize}

\usepackage{graphicx}	
\usepackage{amsmath}	
\usepackage{amssymb}	

\newcommand\eg{{\it e.g.} }

\newcommand\ie{{\it i.e.} }

\newcommand\llrs{long-lived radioactive nuclides }
\newcommand\llrsfullstop{long-lived radioactive nuclides. }
\newcommand\llrscomma{long-lived radioactive nuclides, }
\newcommand\llrsbracket{long-lived radioactive nuclides) }

\title[ ]{ Secondary gas in debris discs released following the decay of long-lived radioactive nuclides, catastrophic or resurfacing collisions}

\author[]{
Amy Bonsor$^{1}$\thanks{E-mail: abonsor@ast.cam.ac.uk}
Mark C. Wyatt$^{1}$,
Sebastian Marino$^{2}$,
Bj\"{o}rn J. R. Davidsson$^{3}$ \\ \newauthor 
  Quentin Kral$^{4}$ and Philippe Thebault$^{4}$
\\
$^{1}$Institute of Astronomy, University of Cambridge, Madingley Road, Cambridge, CB3 0HA, UK\\
$^{2}$ Department of Physics and Astronomy, University of Exeter, Stocker Road, Exeter, EX4 4QL, UK \\
$^{3}$ Jet Propulsion Laboratory, California Institute of Technology, M/S 183-601, 4800 Oak Grove Drive, Pasadena, CA 91109, USA\\ 
$^{4}$ LESIA, Observatoire de Paris, Université PSL, CNRS, Sorbonne Université, Univ. Paris Diderot, Sorbonne Paris Cité, \\ 5 place Jules Janssen, 92195 Meudon, France
}

\date{Accepted 21st September 2023. Received YYY; in original form ZZZ}

\pubyear{2023}

\begin{document}
\label{firstpage}
\pagerange{\pageref{firstpage}--\pageref{lastpage}}
\maketitle

\begin{abstract}

Kuiper-like belts of planetesimals orbiting stars other than the Sun are most commonly detected from the thermal emission of small dust produced in collisions. Emission from gas, most notably CO, highlights the cometary nature of these planetesimals. Here we present models for the release of gas from comet-like bodies in these belts, both due to their thermophysical evolution, most notably the decay of long-lived radioactive nuclides and collisional evolution, including catastrophic and gentler resurfacing collisions. We show that the rate of gas release is not proportional to the rate of dust release, if non-catastrophic collisions or thermal evolution dominate the release of CO gas. In this case, care must be taken when inferring the composition of comets. Non-catastrophic collisions dominate the gas production at earlier times than catastrophic collisions, depending on the properties of the planetesimal belt. We highlight the importance of the thermal evolution of comets, including crucially the decay of long-lived radioactive nuclides, as a source of CO gas around young ($<50$Myr) planetary systems, if large (10-100s kms) planetesimals are present. 

\end{abstract}




\section{Introduction} 

Belts of comets or asteroids, similar to the Kuiper belt, are detected around stars other than the Sun, from the thermal emission of small dust produced in collisions between the larger planetesimals \citep[\eg][]{wyattreview}. The volatile-rich, cometary, nature of the planetesimals in many of these systems is witnessed by the detection of gaseous emission, most commonly CO \citep{Hughes_revew}. This gas has now been detected for tens of planetary systems around AFGM-type stars \citep[\eg][]{Dent2014, Marino2016,Lieman-Sifry2016, Moor2017, Kral2020}. In most cases where the spatial distribution of the gas is resolved, it is associated with the position of the dust belts \citep[\eg][]{Matra2017_betapic}.

Whilst some planetary systems with high masses of CO could be primordial, left-over from the proto-planetary disc phase \citep{Kospal2013}, a secondary origin can readily explain the CO in the many gas-poor debris systems \citep{Kral2017}, as well as in some systems with higher levels of CO \citep[\eg][]{Kral2019}. In other words, the observed gas is not hydrogen-rich gas leftover from the gas-rich protoplanetary disc, but hydrogen-poor gas released from planetesimals in gas-poor debris discs. The CO in this hydrogen-poor gas would only survive on short timescales ($\sim 130$ yrs) due to UV photodissociation \citep{Visser2009}. Instead it is best explained by secondary gas released from the (icy) comets that also produce the dusty material observed in the infrared \citep{ZuckermanSong2012, Dent2014}. Mechanisms suggested for the release of gas include UV desorption \citep{Grigorieva2007}, collisions \citep{Czechowski2007}, radiogenic heating \citep{Davidsson2021} and heating from stellar irradiation \citep{Kral2021}.

In the Solar System, activity triggered by the increased stellar radiation as comets are scattered close to the Sun leads to a cometary tail, with a wide range of species detected including H$_2$O, CO$_2$, CO, CH$_3$OH, H$_2$CO, HCN, H$_2$S and CS$_2$ \citep{Cochran2015}. Cometary tails are dominated by water, whilst a few show spectra dominated by hyper-volatile species including N$_2$ \citep[][\eg comet C/2016 R2]{Biver2018}). The recent New Horizons fly-by of Arrokoth placed upper limits on the release of hyper-volatiles (CO, N$_2$ and CH$_4$) \citep{Gladstone2022}, whilst CO is regularly detected in comets in the inner Solar System \citep{Mumma_Charnley2011}. If the Kuiper-belt is still releasing CO gas today, it is likely too small to be detectable even with in-situ missions like New Horizons \citep{Kral2021}. 
The survival of CO as CO ice in Solar System comets following exposure to stellar irradiation or the decay of long-lived radioactive nuclides including $^{40}$K, $^{232}$Th, $^{235}$U and $^{238}$U is unclear and the CO is potentially trapped in alternate reservoirs including amorphous water ice and CO$_2$ ice as suggested by \cite{Lisse2022, Davidsson2021, Gasc2017, Prialnik1987} and references therein.


This work considers the potential mechanisms that lead to the production of gas in exo-planetary systems, within debris, or cometary belts. The aim is to quantify the gas production rates, such that the conditions (timescales) on which the different mechanisms dominate can be considered. The early release of hyper-volatiles (CO) due to thermophysical evolution driven by heating from the decay of long-lived radioactive nuclides will be compared to the release of hypervolatiles from collisions, both catastrophic and non-catastrophic. Here we focus on the dramatic resurfacing collisions that occur more frequently than catastrophic collisions for those large planetary bodies that are held together by their gravitational strength. Many of the conclusions, however, apply to less violent cratering collisions. Whilst comets likely form during the gas disc phase and the evolution here plays an important role \citep[\eg][]{Simon2022}, this work starts with fully formed planetesimals in a gas-poor environment, akin to a fully formed planetary system.

 This work compares two models for the release of volatiles, the first due to collisions and the second due to radiogenic heating. Whilst in a realistic system both processes might act together, the purpose here is to compare the two processes. The work starts in \S\ref{sec:collisions} by considering collisions, firstly presenting the properties of this planetesimal belt (\S\ref{sec:belt}).  This is followed by details of the collision model for the evolution of the planetesimal belt and the release of volatiles in \S\ref{sec:collmodel}, including mass conservation (\S\ref{sec:mass_conservation}) and the release of CO following collisions (\S\ref{sec:release_CO}). Results from the numerical model for the gas production from planetesimal belts due to collisions are presented in \S\ref{sec:results_collevol}. This is followed by the presentation of a simple model for the volatile release due to radiogenic heating (\S\ref{sec:toy_cometary}) and results for the gas production are compared to those from collisions in \S\ref{sec:results_early}. The validity of the model is discussed in (\S\ref{sec:discussion_validity}), followed by a discussion of whether (and when) radiogenic heating or collisions dominate the gas production in debris belts (\S\ref{sec:discussion_activity_collisions}), the importance of resurfacing collisions, compared to catastrophic collisions (\S\ref{sec:discussion_resurf_cat}), how the observed gas production can provide insights regarding the size of the largest planetesimals present in debris belts (\S\ref{sec:discussion_howlarge}) and the composition of comets, as derived from gas detection (\S\ref{sec:discussion_composition}). The paper is concluded in \S\ref{sec:conclusions}.

\section{Volatile Release from Collisions}
\label{sec:collisions}

The aim here is to quantify the potential release of volatiles, notably CO, in collisions, for comparison with the release from thermophysical evolution. This work considers that many exoplanetary systems have belts of comets or asteroids similar to the Solar System; the key difference being that these planetesimal belts may occur at any radial location and contain significantly more material, as consistent with observations of bright debris discs. In this work, the term planetesimal will be used to refer to the small planetary bodies that are part of the planetesimal belt, regardless of their volatile content and size.

Crucially here we do not assume that the release of volatiles follows the dust evolution of a collisional planetesimal belt, but also consider that volatiles are released in both catastrophic and non-catastrophic collisions. Whilst cratering collisions may play a notable role in the release of CO gas (see discussion in \S\ref{sec:discussion_validity}), we focus here on resurfacing or shattering collisions. That is, the collisions that regularly occur to the largest planetesimals, where sufficient energy is imparted to shatter, but not to overcome self-gravity and disperse the fragments, leading to the formation of a rubble pile. Here we present a framework to follow the collisional evolution of a debris belt, alongside a potential model for how volatiles are released in collisions. Given uncertainties in exactly how volatiles are released in collisions, the model is presented in such a manner that it could be readily adapted to an updated model for the collisional production of gas.  Full details of the variables used in this work can be found in Table~\ref{tab:variables}.

\subsection{Properties of the planetesimal belt}
\label{sec:belt}

The planetesimal belt is characterised by a range of properties, including its total initial mass in dust or solids, $m_{\rm s, tot} (0)$, its radial location, $r$ and width $dr$. The properties of the planetesimals themselves are characterised based on their masses, $M_k$, a size distribution, with slope $\alpha$, from a minimum, $M_{\rm min}$ to a maximum, $M_{\rm max}$, planetesimal mass, a density $\rho_k$ and a CO content, or volatile mass fraction, $f_{v,k}$. In this work, the volatile is considered to be CO, but the model could be applied to trace the evolution of any volatile, including water ice, and solids refers to the dusty component of comets.

In order to numerically follow the evolution of the belt, the mass in the planetesimal belt is distributed between bins, labelled by the mass in individual solid planetesimals in each $k$th bin, $M_{\rm s, k}$. The planetesimals in each bin have a total mass which is the sum of the mass in dust or solids, $M_{\rm s,k}$ and volatiles, $M_{\rm v,k}$, such that $M_k = M_{\rm s, k} + M_{\rm v, k}$. The bins are logarithmically spaced in $M_{\rm s,k}$, and $k=1$ labels the bin of largest mass. The logarithmic bin spacing $\delta$ is defined such that $1-\delta= \frac{M_{\rm s,k+1}}{M_{\rm s,k}}$.  The initial number of planetesimals in each mass bin is assumed to scale as: 
 \begin{equation}
 n_k(M)= KM_{\rm s,k}^{-\alpha},
     \label{eq:size_d}
 \end{equation} which is equivalent to the commonly used $n(D)dD\propto D^{-\alpha'}dD$, where $\alpha=(\alpha'+2)/3$ \citep[see \eg][]{wyatt11, Dohnanyi}, such that when $\alpha'$=7/2, $\alpha = 11/6$.  Thus, the total mass of solid planetesimals in each mass bin, is given by integrating the mass in solids across the bin between $M_{\rm s,k}$ and $M_{\rm s, k}(1-\delta)$ to give: 
  \begin{equation}
     m_{s, k} (0)= K M_{\rm s, k}^{2-\alpha},
     \label{eq:m_s}
 \end{equation}
 assuming $\delta <<1$, where  $K= \frac{{m_{\rm s, tot}(0)}}{\Sigma_{i=1}^{i_{\rm max}} M_{{\rm s}, i}^{2-\alpha}}$ and ${i_{\rm max}}$ labels the bin of the smallest planetesimal present in the collisional cascade.


The planetesimals in the $k$th bin have an average density, $\rho_k$ and diameter $D_k$, where $M_{k} = \frac{\pi \rho_k D_k^3}{6} $. The diameter of the planetesimal, $D_k$ in each bin, as well as their average density, $\rho_k$, can change as a function of time as the planetesimals lose volatiles, whilst the mass in solids, $M_{\rm s, k}$ remains constant. Thus, the volatile fraction of each planetesimal, $f_{{\rm v},k}$, also changes as a function of time and is denoted:
\begin{equation}
f_{{\rm v},k}= \frac{M_{{\rm v},k}}{M_{{\rm s},k}+M_{{\rm v},k}}.
\end{equation}
The average density is given by: 
\begin{equation}
\rho_k= \frac{1}{((1-f_{{\rm v},k})/\rho_s + f_{{\rm v},k} /\rho_{\rm v})},
\label{eq:rho}
\end{equation}
 where $\rho_s$ is the density of the dust or solid component and $\rho_{\rm v}$ is the density of the volatile component. The number of colliders in the $k$-th bin is given by the ratio of the total mass in solids to the mass of each planetesimal in solids, as the total diameter or mass in volatiles in the $k$-th bin may change, $n_k=\frac{m_{s,k}}{M_{s,k}}$.

\subsection{Conditions for catastrophic/resurfacing collisions} 
\begin{figure}
\includegraphics[width=0.48\textwidth]{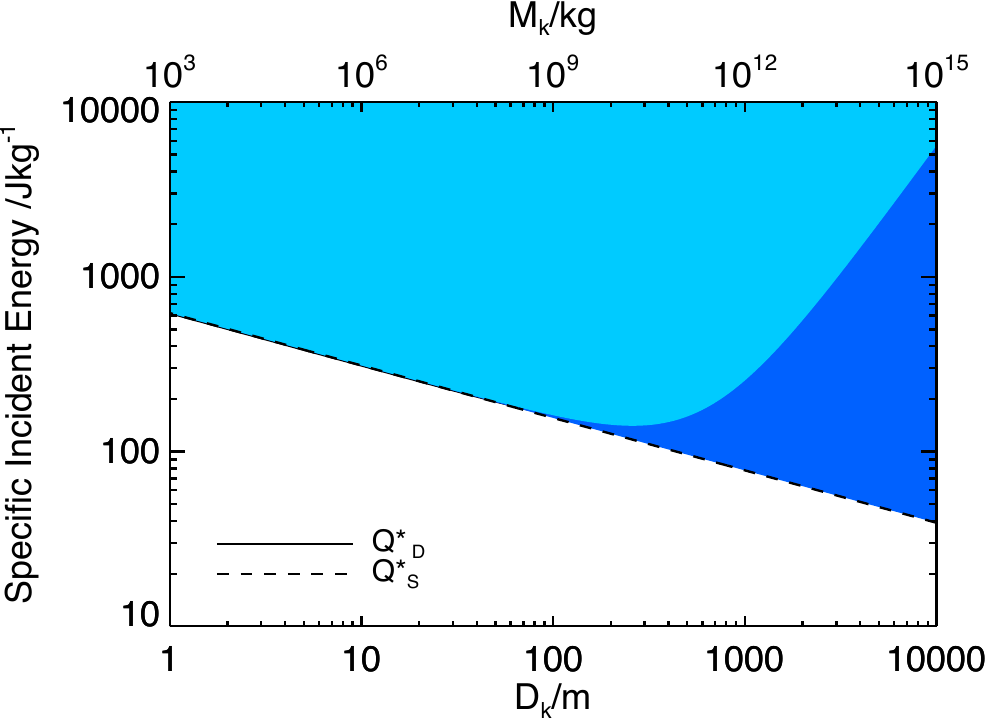}

\caption{The dispersal threshold, $Q_D^*$ (solid line) and shattering threshold, $Q_S^*$ (dashed line), as a function of the planetesimal's total diameter, $D_k$ or mass $M_k$, calculated using the parameterised form of Eq.~\ref{eq:qdstar} and Eq.~\ref{eq:qsstar} from \citet{benzaphaug,Durda1998}. The light-blue shaded region indicates those collisions that will be catastrophic, whilst the darker blue region those collisions that are shattering.  }
\label{fig:qdstar}
\end{figure}

Catastrophic collisions are those with sufficient energy to disrupt a planetesimal, leaving no remnant larger than half the mass of the original planetesimal. The incident energy must be above the specific incident (impact) energy required to cause a catastrophic collision, or the dispersal threshold, $Q_D^*$. A power-law form for the dispersal threshold is assumed, following work on collision outcomes by \cite{benzaphaug,Durda1998},  such that: 
\begin{equation}
Q_D^*=Q_a D^{-a} + Q_b D^b,
\label{eq:qdstar}
\end{equation}
where $a$ and $b$ are both positive constants related to the planetesimal's material and gravitational strength, respectively. Following \cite{wyatt11}, we take $Q_a = 620$ J kg$^{-1}$, 
$a = 0.3$, $Q_b=5.6\times 10^{-3}$ J kg$^{-1}$, 
and $b = 1.5$, where $D$ is in m, noting that these parameters are applicable to basalt rather than a mixture of ice and basalt. This is plotted in Fig.~\ref{fig:qdstar} as a function of the planetesimal's total diameter, $D_k$ or mass, $M_k$, assuming an average density of $\rho_k=1$g$\,$cm$^{-3}$. For small planetesimals, the dispersal threshold is dominated by the planetesimal's material strength, whereas for larger planetesimals, gravity dominates. 

Resurfacing collisions occur when there is sufficient energy to disrupt a planetesimal, but insufficient energy to disperse the fragments subsequently. 
Resurfacing collisions occur when the incident energy is above the specific incident energy required for shattering, $Q_{S}^*$, but below the specific incident energy required for dispersion, $Q_D^*$, where: 
\begin{equation}
Q_S^*=Q_a D^{-a}.
\label{eq:qsstar}
\end{equation}
This shattering threshold is plotted, alongside the dispersal threshold, on Fig.~\ref{fig:qdstar}. The minimum in the dispersal threhold occurs for diameters of size $D_W$, where 
\begin{equation}
D_W =\left(\frac{aQ_a}{bQ_b}\right)^{\frac{1}{(a+b)}}.
\label{eq:dw}
\end{equation}
Resurfacing collisions can occur for all sizes, however, the incident energy range for which $Q_D^*>Q>Q_S^*$, the condition required for a resurfacing collision to occur, becomes vanishingly small for small diameter particles. In the numerical model a cut-off is introduced, where no resurfacing collisions are assumed to occur if there are less than three mass bins between minimum mass bin above which resurfacing collisions would occur, labelled $i_{\rm rk}$ and the minimum mass above which catastrophic collisions would occur, labelled $i_{\rm ck}$. Or in other words, no resurfacing collisions occur for $M_{{\rm min}, r}$. This avoids resurfacing collisions from switching on and off around diameter bins of a few hundred kms (depending on the properties of the planetesimal belt). For these bodies, the smallest bodies that have sufficient energy to shatter a planetesimal of mass, $M_{\rm min}, r$, also have sufficient energy to catastrophically destroy it.

\begin{figure}
    \centering
    \includegraphics[width=0.48\textwidth]{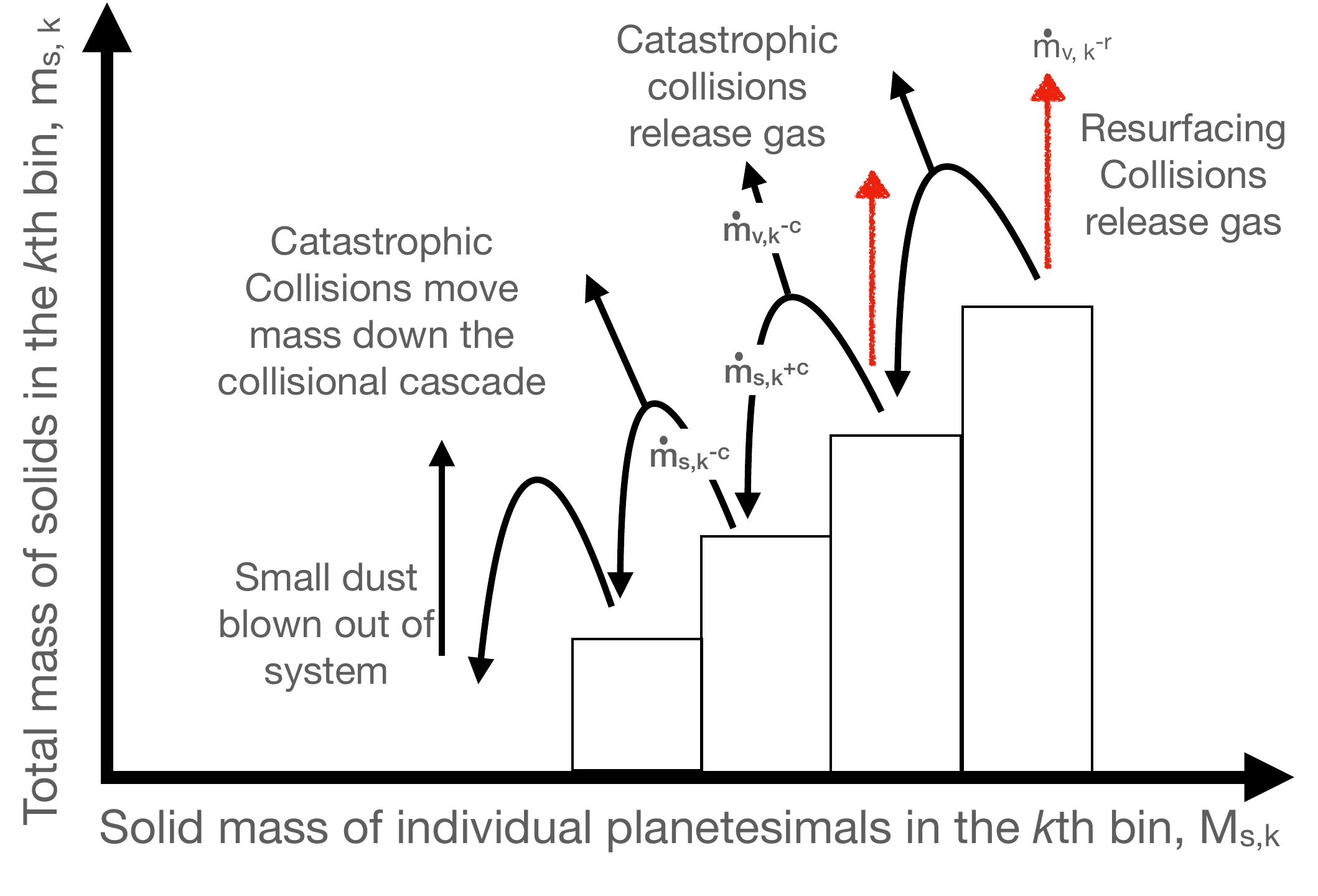}
    \caption{ A cartoon to illustrate how the movement of material following collisions is traced in the model presented in \S\ref{sec:collmodel}. +c refers to the mass gained from catastrophic collisions, -c to that lost from catastrophic collisions, shown by the black arrows and -r to the mass lost from resurfacing collisions, shown by the red arrows  } 
    \label{fig:di_sizedistribution}
\end{figure}

\subsection{Collision Model}
\label{sec:collmodel}
In order to model the collisional evolution of the planetesimal belt, we follow \cite{wyatt11}, with the additional ability to trace both solids and volatiles. A numerical method is utilised. The total mass in each bin is traced at each timestep, $\delta_t$. A particle-in-a-box approach is used to model the rate of collisions between planetesimals in different mass bins, tracing both catastrophic and re-surfacing collisions. Following each catastrophic collision, mass is redistributed amongst smaller fragments, until it eventually becomes sufficiently small to be blown out of the system by radiation pressure (see Fig.~\ref{fig:di_sizedistribution}). In such a manner, the planetesimal belt is depleted in mass as a function of time. Additionally to \cite{wyatt11}, the model presented here traces the volatile content of planetesimals separately to their refractory (solid) content. In other words, the total mass in solids in the $k$-th bin is given by $m_{\rm s, k}$, whilst the total mass in volatiles is given by $m_{\rm v, k}$. At each timestep, volatiles can be released to gas and the total mass in gas is also traced $m_{\rm gas}$. A full list of the variables used in this model can be found in Table~\ref{tab:variables}.

\subsubsection{Collision Rates} 
The rate of collisions between particles in the $k$-th bin, with particles in the $i$-th bin is determined by the cross-sectional area for collisions, $\frac{\pi}{4}(D_k + D_i)^2$, the relative velocity, $v_{\rm rel}$, and the volume through which the planetesimals are moving, $V$. Planetesimals in the $k$-th bin can only collide catastrophically with particles larger than a certain size, which corresponds to those planetesimals in size bins with indices less than $i_{ck}$, such that:  
\begin{equation}
R_k^{c}= \Sigma_{i=1}^{i_{ck}} \frac{n_i}{4} (D_k + D_i)^2 \frac{ \pi v_{\rm rel}}{V},
\label{eq:rate_c}
\end{equation}
where $n_i$ is the number of colliders in the $i$-th bin and $V$ is the volume through which the planetesimals are moving, given by \citep{Sykes1990}:
\begin{equation}
    V = 8 \pi r^3 e \sin (I) (1+ \frac{e^2}{3}),
    \label{eq:v}
\end{equation} 
where $r$ is the belt radius, $e$ the average eccentricity and $I$ the maximum inclination of particles. The index $i_{ck}$ labels the bin containing planetesimals of mass, $M_{\rm ck}$, the smallest bodies that can cause catastrophic collisions to planetesimals in the $k$-th bin, where:  
\begin{equation}
     M_{ck}=\left(   \frac {2 Q_{D}^*} {v_{\rm rel}^2} \right)  M_k.  
     \label{eq:mck}
\end{equation}
We note here that if the minimum mass that can catastrophically destroy planetesimals is larger than the size of the planetesimals, \ie $M_{\rm ck}>M_k$ the premise of the model breaks down. This is because in such a simple model, the largest planetesimals would no longer evolve collisionally, whilst in reality they would lose mass due to e.g. cratering collisions. We note that the approximations break down for targets larger than 30km ($D>30$km) for a belt at 100au, with $v_{\rm rel}= e \, v_k$, where $e=0.1$ and the model for $Q_D^*$ used here.

The rate of resurfacing collisions is calculated in a similar manner, with only colliders too small to result in catastrophic collisions ($i<i_{\rm ck}$) and sufficiently large to cause resurfacing collisions, $i_{rk}$ considered:
\begin{equation}
R_k^{r}= \Sigma_{i=i_{ck}}^{i_{rk}} \frac{n_i}{4} (D_k + D_i)^2 \frac{ \pi v_{\rm rel}}{V},
\label{eq:rate_r}
\end{equation}
where $i_{rk}$ labels the bin of mass $M_{rk}$ that contains the smallest impactors that can cause resurfacing collisions, where: 
\begin{equation}
    M_{rk}=\left(   \frac {2 Q_{S}^*} {v_{\rm rel}^2} \right)  M_k.
    \label{eq:mrk}
    \end{equation}


The average lifetime of a planetesimal of diameter, $D$, against catastrophic collisions can be calculated as $t_c = \frac{1}{R_k^c}$ (Eq.~\ref{eq:rate_c}), and similarly for resurfacing collisions as $t_r=  \frac{1}{R_k^r}$ (Eq.~\ref{eq:rate_r}).

\begin{figure}
\includegraphics[width=0.48\textwidth]{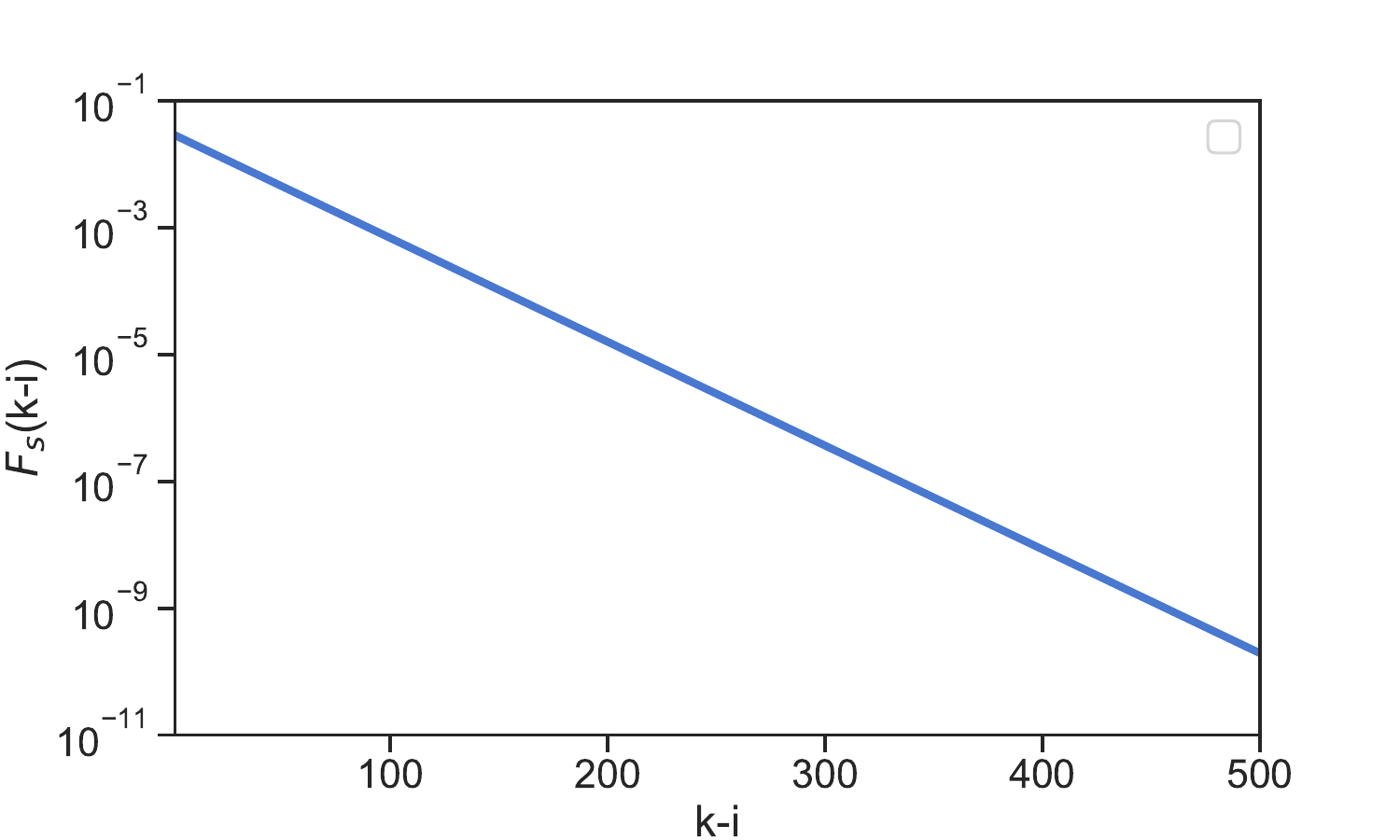}

\caption{The fraction of mass from collisions in the $i$-th bin that ends up in the $k$-th bin, or the redistribution function, for $\delta=0.01$ and $\alpha=11/6$.   }
\label{fig:redistribution}
\end{figure}

\subsection{Mass Conservation}
\label{sec:mass_conservation}

The mass in solids in the collisional cascade is depleted with time as catastrophic collisions grind down the planetesimals into dust that is lost from the system. The mass in volatiles can be lost to gas from planetesimals of all sizes. However, at every timestep the total mass is conserved, such that the rate of change in the $k$th bin of the total mass in solids (volatiles) $\dot{m}_{{\rm s},k}$ ($\dot{m}_{{\rm v},k}$) is given by:  
\begin{eqnarray}
{\dot m}_{{\rm s},k} &=&  \dot{m}_{{\rm s},k}^{+c} - \dot{m}_{{\rm s},k}^{-c}, \label{eq:mdot_c}\\
\dot{m}_{{\rm v},k} &=& \dot{m}_{{\rm v},k}^{+c} - \dot{m}_{{\rm v},k}^{-c} -  \dot{m}_{{\rm v},k}^{-r}, \label{eq:mdot_v} \\
\dot{m}_{\rm g} &=& \Sigma_{k=1}^{k_{\rm max}} (\dot{m}_{{\rm g},k}^{+c} +\dot{m}_{{\rm v},k}^{-r})  + \Sigma_{k=k_{\rm max}}^{\infty}\dot{m}_{{\rm g},k}^{+c}\label{eq:mgas} ,
\label{eq:massconservation}
\end{eqnarray}
where $\dot{m}_{{\rm s},k}^{-c}$ ($\dot{m}_{{\rm v},k}^{-c}$) is the rate at which the total mass in solids (or volatiiles) in the $k$-th bin is lost to catastrophic collisions, $\dot{m}_{{\rm s},k}^{+c}$ is the rate at which the mass in solids is gained from catastrophic collisions of larger bodies, $\dot{m}_{{\rm v},k}^{+c}$ is the rate at which mass in volatiles is gained from catastrophic collisions of larger bodies, noting that this accounts for the volatiles mass lost to gas, $\dot{m}_{{\rm g},k}^{+c}$ is the mass lost to gas directly as the $k$-th bin gains mass in volatiles from catastrophic collisions in larger bins and $\dot{m}_{{\rm v},k}^{-r}$ is the rate of mass loss in volatiles from re-surfacing collisions. The mass in gas, $m_{{\rm g}}$, is increased by mass loss from volatiles to gas in both catastrophic and re-surfacing collisions, from the material received in the $k$th bin at a rate of $\dot{m}_{{\rm g},k}^{+c}$ and $\dot{m}_{{\rm v},k}^{+r}$, respectively. Additionally, some gas is produced when volatile fragments directly to particles smaller than the minimum present in the collisional cascade, which leads to the additional term, $\Sigma_{k=k_{\rm max}}^{\infty}\dot{m}_{{\rm g},k}^{+c}$.

The mass loss rate for solids is given by 
\begin{equation}
\dot{m}_{{\rm s},k}^{-c}= m_{{\rm s},k} R_k^{c}, 
\end{equation}

whilst that for volatiles, exclusively due to catastrophic collisions, is given by 
\begin{equation}
\dot{m}_{{\rm v},k}^{-c}= m_{{\rm v},k} R_k^{c}.
\end{equation}
 
The mass rate gained for solids is given by 
\begin{equation}
\dot{m}_{{\rm s},k}^{+c}= \Sigma^{i_{mk}}_{i=1} F(k-i) \; \dot{m}_{{\rm s},i}^{-c}, 
\end{equation} 

where $F(k-i)$ is the fraction of the mass leaving the $i$-th bin from collisions that goes into the $k$-th bin, or the redistribution function, which we assume to be scale independent. We assume that fragments produced in catastrophic collisions have a range of masses from the largest fragment, with $\frac{M_{{\rm s},i}}{2}$, which falls in the bin labelled $i_{lr}$, to the smallest body considered, which falls in the bin labelled by $i_{max}$, which we assume to be much smaller than $\frac{M_{{\rm s}, i}}{2}$. This is a good approximation for collisions destroying bodies with $D\gg$ cm as the smallest particles in the disc will be mm-sized or smaller. Thus, the $k$-th bin can only gain mass from catastrophic collisions between objects with a mass $2M_{{\rm s}, k}$ or greater, labelled by $i_{mk}=k  - \frac{ln (2)}{\delta}$. Thus, the mass rate gained for solids in the $k$-th bin is calculated by summing over the contributions from the largest mass bin, $i=1$, down to $i_{mk}$, which labels the bin of mass $2M_{{\rm s}, i}$. We assume that the size distribution of fragments is given by Eq.~\ref{eq:size_d}, where $\alpha>1$ and the separation between bins $\delta \ll 1$. We consider the fragmentation of a body that is a uniform mixture of volatiles (ices) and solids, and that, therefore, all the fragments retain the same uniform mixture. This leads to a redistribution function, where we assume the same exponent for the power-law, $\alpha$, as the size distribution, given by :
 \begin{equation}
     F_s(k-i) = \eta^{(\alpha-1)} (1-\delta) ^{(k-i)(1-\alpha)} \, \delta (1-\alpha) 
     \label{eq:f_s}
 \end{equation}

This is based on Eq.~20 of \cite{wyatt11}, where $\delta$ is now the spacing between mass bins and not radial bins, $\eta_{\rm max}=1/2$, such that $\alpha'=3 \alpha-2$, where $\alpha'$ are the parameters used in \cite{wyatt11} and $\delta \sim 3 \delta'^3 $. 
This function plotted in Fig.~\ref{fig:redistribution}, truncated at $(k-i)=\frac{ln (2)}{\delta}=69$, which for $\delta=0.01$ labels the bin of solid mass, $M_{{\rm s}, i}/2$, or the largest fragment of a catastrophic collision. By definition all of the mass leaving the $i$-th bin ends up in bins between $k=i_{\rm lr}$ and $k=\infty$, such that 
\begin{equation}
\Sigma_{k=i_{\rm lr}}^{\infty} F(k-i)=1.
\end{equation}

In a similar manner, the rate of mass gain for volatiles is given by : 
\begin{equation}
\dot{m}_{{\rm v},k}^{+c}= \Sigma_{i=1}^{i_{mk}} F_{\rm v}(k-i, f_{ {\rm v}, i}) \dot{m}_{{\rm v},i}^{-c} ,
\end{equation}
where the $F_{\rm v}(k-i, f_{ {\rm v}, i})$ is the redistribution function for volatiles. This is explicitly a function of the volatile fraction of the disrupting bodies, $f_{ {\rm v}, i}$, as this accounts for the possibility that some of these volatiles may be lost soon after the collision.

This is related to the redistribution function for solids, in that:
\begin{equation}
F_{\rm v}(k-i, f_{ {\rm v}, i})= F_s(k-i) \, (1-\chi_k^c(f_{{\rm v}, i})),
\end{equation}
where $\chi_k^c(f_{{\rm v}, i})$ is the fraction of the volatile mass lost to gas as soon as the fragment is created, calculated in the following section, which is a function of the total mass of the newly formed body, $M_k$, and the volatile fraction of the original planetesimal in the $i$-th bin, $f_{{\rm v},i}$. 

In a similar manner, a fraction $\chi_k^c(f_{{\rm v}, i})$ of the volatiles in the fragment gained by the $k$-th bin from catastrophic collisions is released directly to gas, from catastrophic collisions between larger bodies, is released directly to gas. The rate at which this occurs in the $k$-th bin is given by: 
\begin{equation}
\dot{m}_{{\rm g},k}^{+c}= \Sigma_{i=1}^{i_{mk}}F_{\rm s}(k-i) \, \chi_k(f_{{\rm v}, i}) \,\dot{m}_{{\rm v},i}^{-c}. 
\label{eq:mgk_c}
\end{equation}

The rate of mass lost for volatiles, due to re-surfacing collisions, is given by:
\begin{equation}
\dot{m}_{{\rm v},k}^{-r}= \chi_{k}^r(f_{v,k}) m_{{\rm v},k} R_k^{r}. 
\end{equation}

\subsection{Release of CO following collisions } 
\label{sec:release_CO}

The exact role of collisions in releasing volatiles from cometary bodies is not clear. Experimental work is best at probing collisions between small particles \citep{BlumWurm2008, Simon2022} and tracking of volatile species is limited. In the Solar System, whilst activity and degassing in individual active comets can be monitored, linking this to a comet's collision history is challenging. Nonetheless, collisions will clearly play a role in releasing volatiles from comets. Collisions can transfer heat to planetesimals, which in turn leads to volatile loss \citep[\eg][]{Jutzi2017, Davidsson2023}. Collisions also expose new surface, such that volatiles can be lost via sublimation or UV desorption. Rather than attempting to model these processes in detail, here we produce a simple model in which we consider the efficiency at which volatiles are lost in an individual collision to be a function of the surface area of the colliders. The model is set up in such a way that this precription could be updated in the future. The model is equivalent to the devolatisation of a thin layer of depth, $h$. The depth of this layer may be relatively large, for example for comets sufficiently close to the star that sublimation acts on long timescales (tens of metres), but may be very small for example for comets far from the star, if thermal or UV desorption of the fresh surface layer is the only mechanism to release volatiles (less than millimetres), as the rest of the CO would remain trapped. 

\subsubsection{Model for release of volatiles in Catastrophic Collisions}
\label{sec:release_Cat}
We consider a simple model in which following a catastrophic collision the fractional release of volatiles is proportional to the surface area of the fragments produced, or in other words volatiles are released from the equivalent of a surface layer of depth, $h$. Although we acknowledge here that this layer may not truly be a thin surface surrounding the whole comet, but instead be focused around the impact site. We consider the mass in volatiles released to gas due to fragments produced by catastrophic collisions of solid mass, $M_{{\rm s}, k}$, to be given by the mass in volatiles found in the layer, $h$, which can be calculated by subtracting the volatile mass of the smaller planetesimal ($\propto (R-h)^3$) from the total volatile mass of the planetesimal ($\propto R^3)$. Thus, the mass released to gas by fragments entering the $k$-th bin is given by: 
\begin{equation}
\delta M_{g, k}^c =   4 \pi \rho_k f_{v, k}  \left ( \left (\frac{3 M_k}{4\pi \rho}\right)^{2/3}\, h -  \left (\frac{3 M_k}{4\pi \rho}\right)^{1/3}\,h^2 + \frac{h^3}{3} \right)
 \label{eq:delta_mvk}
\end{equation}
such that the fractional release of volatiles, which can never be greater than one, is given by:
\begin{equation}
\label{eq:chik}
   \chi_k^c = \frac{\delta M_{v,k} }{f_{v,k} M_k}. 
\end{equation}

The solid lines on Fig.~\ref{fig:chi_tot} shows the fraction of volatiles arriving in the $k$-th bin from the disruption of larger planetesimals that are released directly to gas, for different assumptions about the depth of the layer, $h$. The fractional release decreases with increasing particle size for small particles, as the assumption of a constant depth, $h$, accounts for a larger fraction of the body. Alternatively, the fractional release to volatiles due to the catastrophic destruction of planetesimals of mass, $M_k$, can be considered as the sum of the mass lost from all fragments produced and is shown by the dotted lines on Fig.~\ref{fig:chi_tot}.

\subsubsection{Model for the release of volatiles in Resurfacing Collisions}
\label{sec:release_resurf}

We consider that a large planetesimal of mass $M_k$, following a resurfacing collision is split into fragments of mass $M_f$ (diameter, $D_f$){, the largest of whom has mass $M_f/2$, which occurs in the bin labelled $i_{\rm frag}$.} Each fragment individually loses volatiles to gas from an outer layer of depth, $h$,  such that the volatile loss, $\delta M_{\rm g, f}$, is given by Eq.~\ref{eq:delta_mvk}.

The mass released to gas from volatiles in a collision of a body of mass, $M_k$ is, thus, the sum of the mass of fragments in each mass bin multiplied by the fraction of that mass released to gas from volatiles $\chi_f$: 
\begin{equation}
   \delta m_{\rm g, k}^r = \frac{f_{\rm v,k}}{(1-f_{\rm v,k})}\,\Sigma_{f=i_{\rm frag}}^{i_{\rm max}} m_{s,f} \, \chi_f,
\end{equation}
where $\chi_f$ comes from Eq.~\ref{eq:chik}, $m_{s,f}= K' \, M_{s,f}^{2-\alpha} $ and the constant of normalisation $K'$ is given by the mass in solids of a planetesimal in the $k$th bin, $(1-f_{\rm v, k}) M_k=K' \, \Sigma_{i=i_{\rm frag}}^{i_{\rm max}}M_{s,i}^{2-\alpha} $. Thus, the fraction of the volatile mass released to gas by a resurfacing collision is given by: 

\begin{equation}
    \chi_{k}^r =  \frac{  \Sigma_{f=i_{\rm frag}}^{i_{\rm max}}  M_{s,f}^{2-\alpha} \chi_f } {\Sigma_{i=i_{\rm frag}}^{i_{\rm max}}M_{s,i}^{2-\alpha} } . 
   \label{eq:chi_kr}
\end{equation}

\begin{figure}
\includegraphics[width=0.48\textwidth]{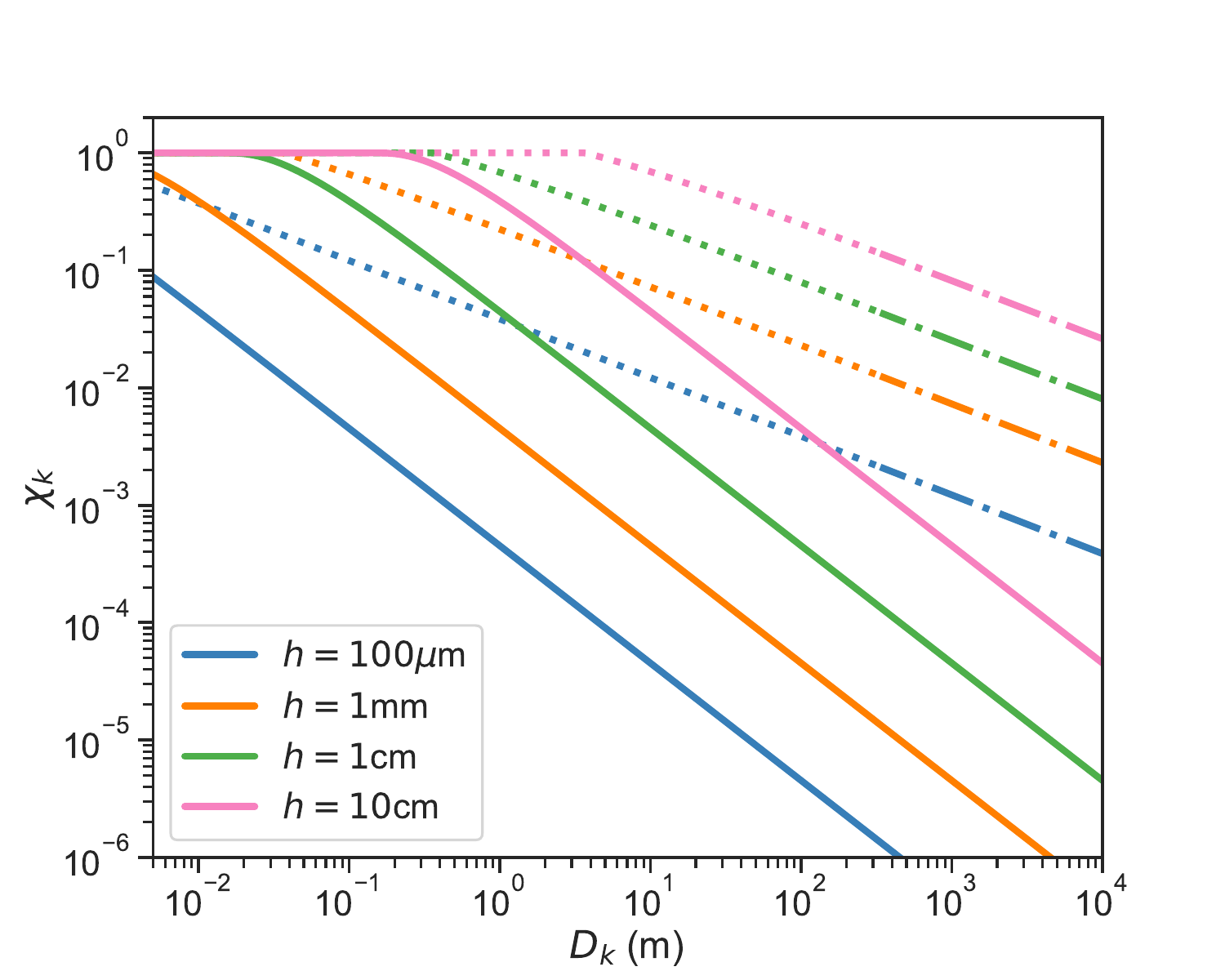}
\caption{The fractional release of volatiles directly to gas following catastrophic or resurfacing collisions, as a function of the fragment mass or diameter, $D_k$,  for different assumptions regarding the thin surface layer of depth $h=100\mu$m, 1mm, 1cm or 10cm from which volatiles are released. The solid lines show the fraction of the volatile mass arriving in the $k$th bin due to catastrophic collisions of larger bodies that is released directly to gas (Eq.~\ref{eq:chik}). The dot-dashed lines show the fractional release of volatiles to gas from resurfacing collisions of bodies of mass, $M_k$ (Eq.~\ref{eq:chi_kr}). The dotted lines, which overlap with the dot-dashed lines, show the fractional release to gas following catastrophic collisions of bodies of mass, $M_k$. }

\label{fig:chi_tot}
\end{figure}

The dot-dashed lines on Fig.~\ref{fig:chi_tot} shows the fractional release of volatiles to gas from resurfacing collisions of planetesimals in the $k$th bin.

\subsection{Numerical Simulations }
\label{sec:numerical}
The CO production from a planetesimal belt is calculated numerically at every timestep by splitting the planetesimal belt into logarithmically spaced solid mass bins of width $\delta$, using the size distribution (Eq.~\ref{eq:size_d}). The dust and gas production from collisions are calculated using Eqs.~\ref{eq:massconservation}, with a numerical method invoked to trace the mass in solids and the mass in volatiles in every bin $m_{{\rm s}, k}$ and $m_{{\rm v}, k}$, as a function of time. This solves 
Eqs.~\ref{eq:massconservation}, tracking the volatiles lost to gas at each timestep. The timestep is selected such that the mass lost by the smallest bin is less than half of the mass initially in the smallest bin. $F_s$ is normalized such that Eq.~\ref{eq:f_s} applies exactly, counting only bins up to a large number $2n_{\rm bin}$, where $n_{\rm bin}$ is the total number of bins used. Mass conservation is assured by tracking the total mass in solids or volatiles at every output and adding the very small additional mass lost to gas or dust. The evolution of the solids in the code is benchmarked against \cite{wyatt11} using an initial mass distribution with $\alpha=0.86$ ($\alpha'=-3.6$), a bin width of $\delta =0.02$, and a belt between 7.5 and 11au to match \cite{wyatt11}.

\begin{figure}

\includegraphics[width=0.48\textwidth]{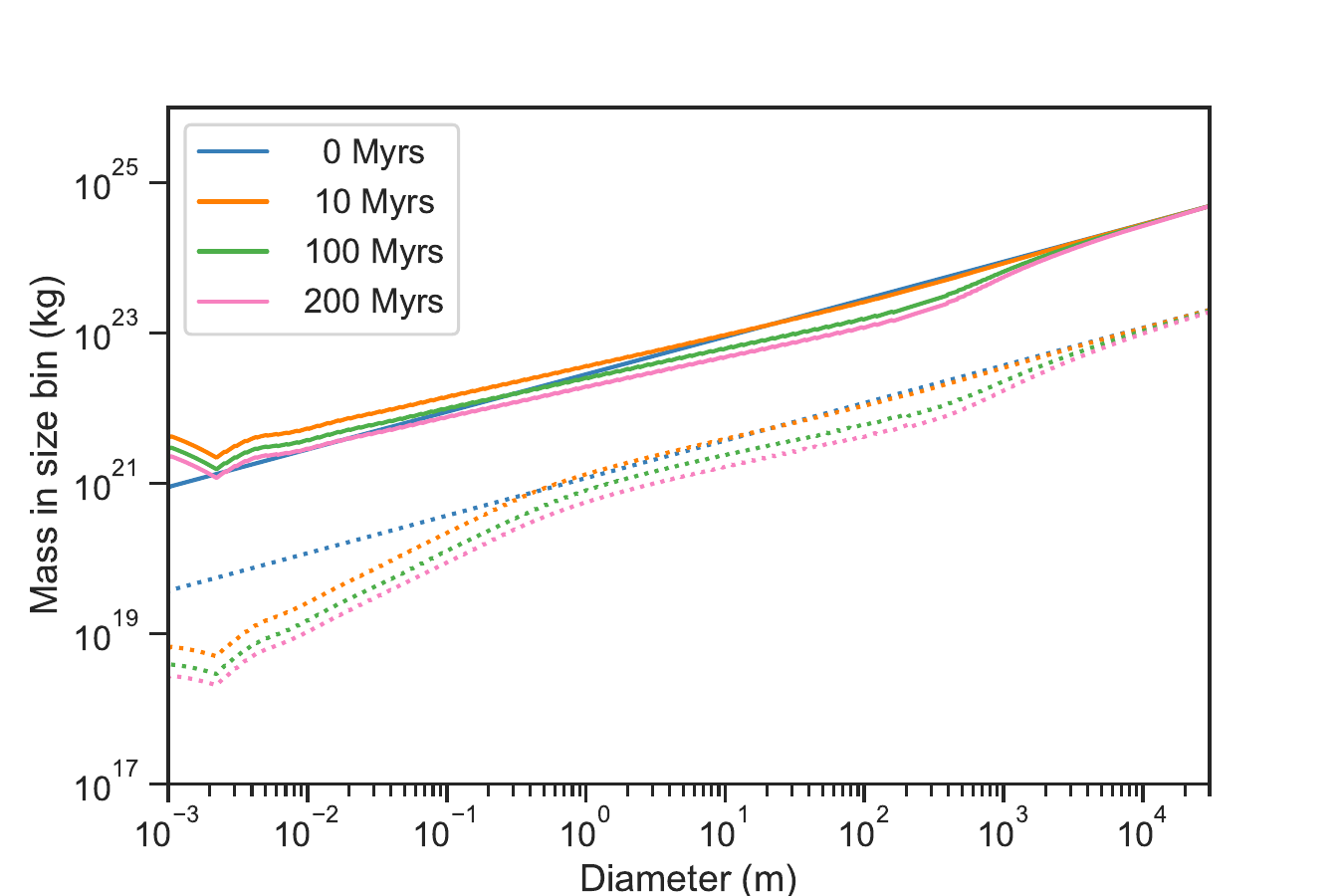}
\includegraphics[width=0.48\textwidth]{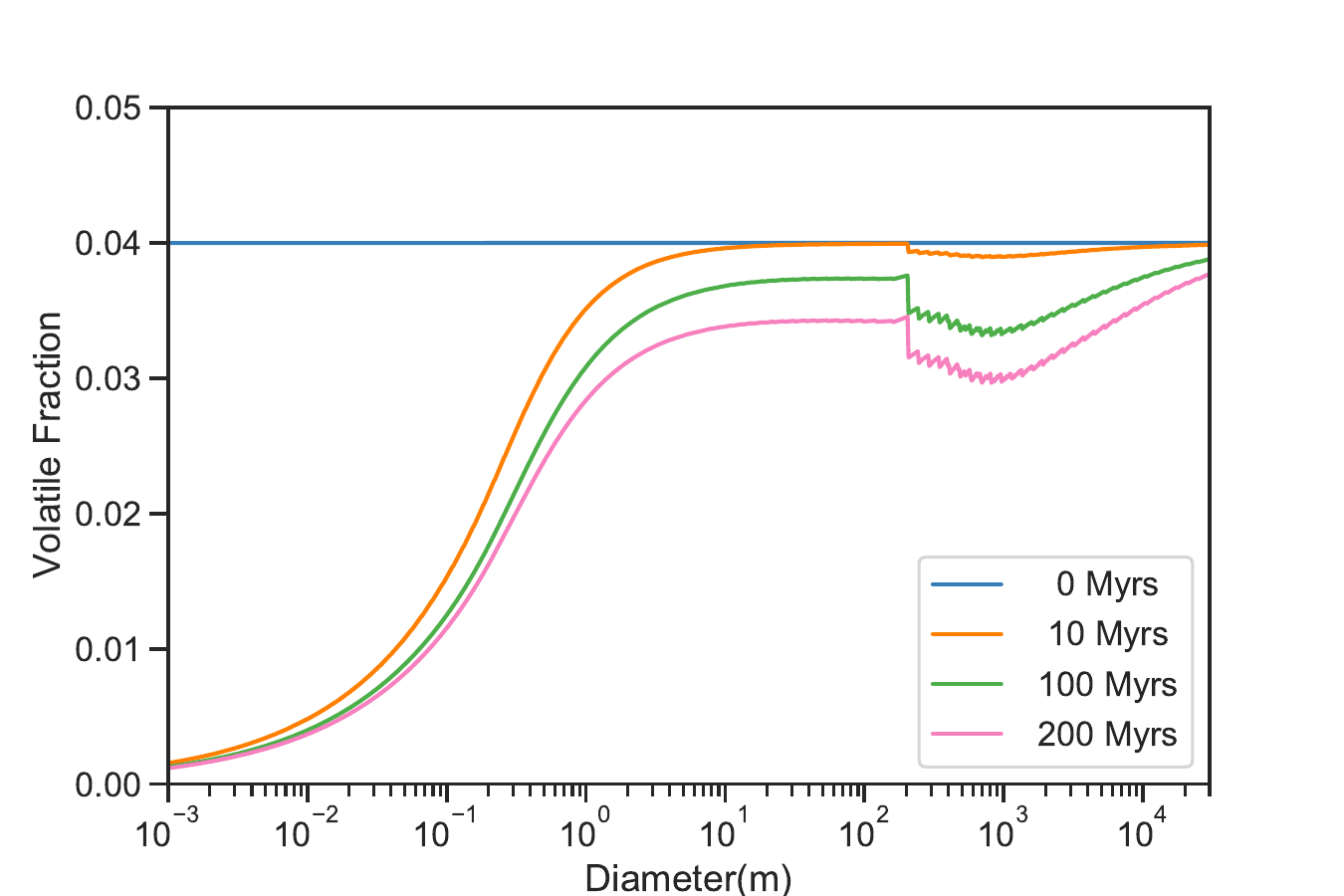}
\includegraphics[width=0.48\textwidth]{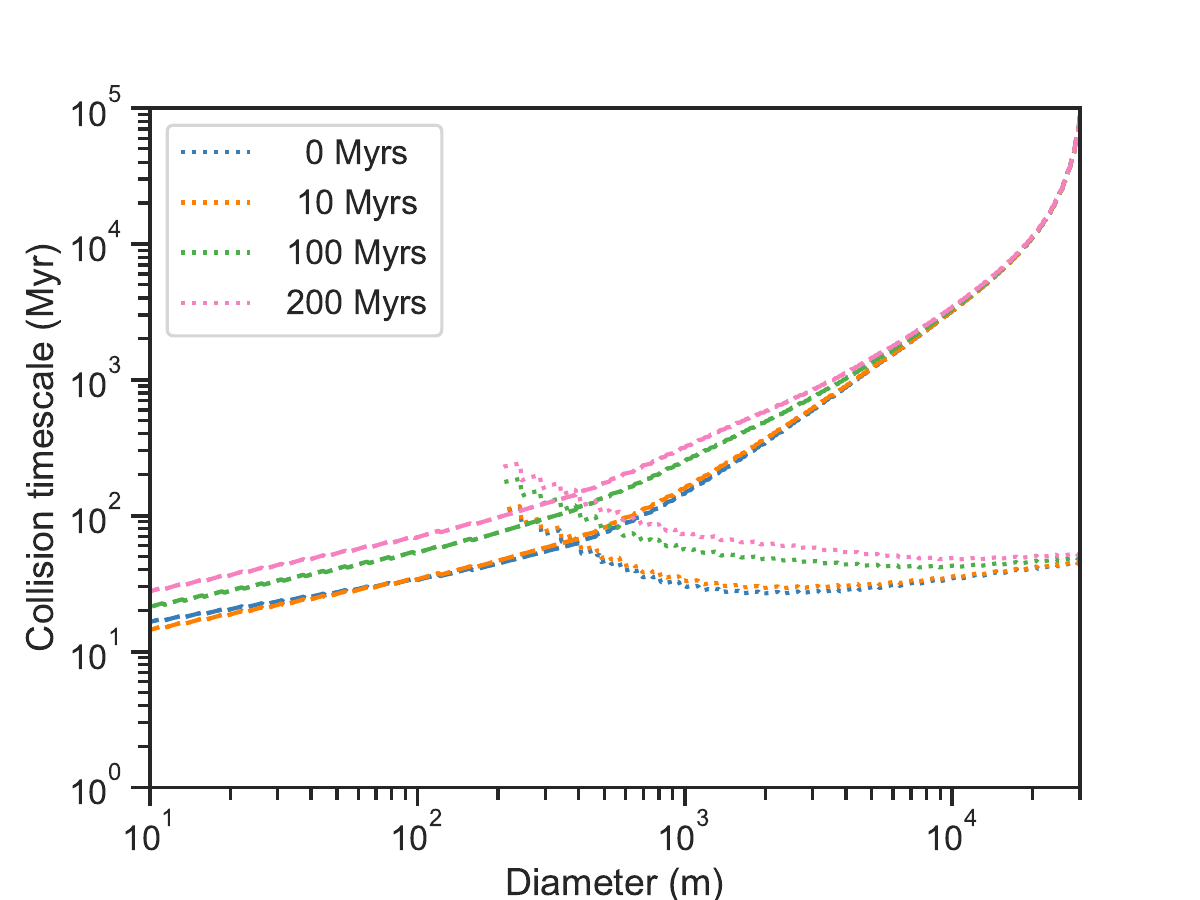}

\caption{The collisional evolution of the fiducial belt at 100au, with an initial total disc mass of 100$M_\oplus$ in particles between 1mm and 30km, with $\alpha=11/6$, an initial volatile content of 4\%,  $\delta=0.05$ and $h=10$cm. The top panel shows the mass in each diameter (mass) bin (size distribution) in both solids (solid lines) and volatiles (dotted lines) at various epochs.  The middle panel shows the volatile content of planetesimals as a function of diameter. The solid line is the initial volatile content of 4\%. Catastrophic collisions erode volatiles in small bodies, whilst resurfacing collisions erode volatiles in the largest planetesimals. The bottom panel shows the timescale for both catastrophic (dashed lines) and resurfacing  (dotted lines) collisions, calculated using the inverse of Eqs.~\ref{eq:rate_c},~\ref{eq:rate_r}. The jagged lines are an artefact of the finite bin size used in the simulations. The sharp transition in volatiles at around $D=200$m is to the unrealistic abrupt onset of resurfacing collisions assumed by the model. }The bump seen at small diameters is due to the finite cut-off at $D_{\rm min}=1$mm, which may be artificial, if a more realistic disc extended down to smaller sizes. 
\label{fig:example}
\end{figure}

\begin{figure}
\includegraphics[width=0.48\textwidth]{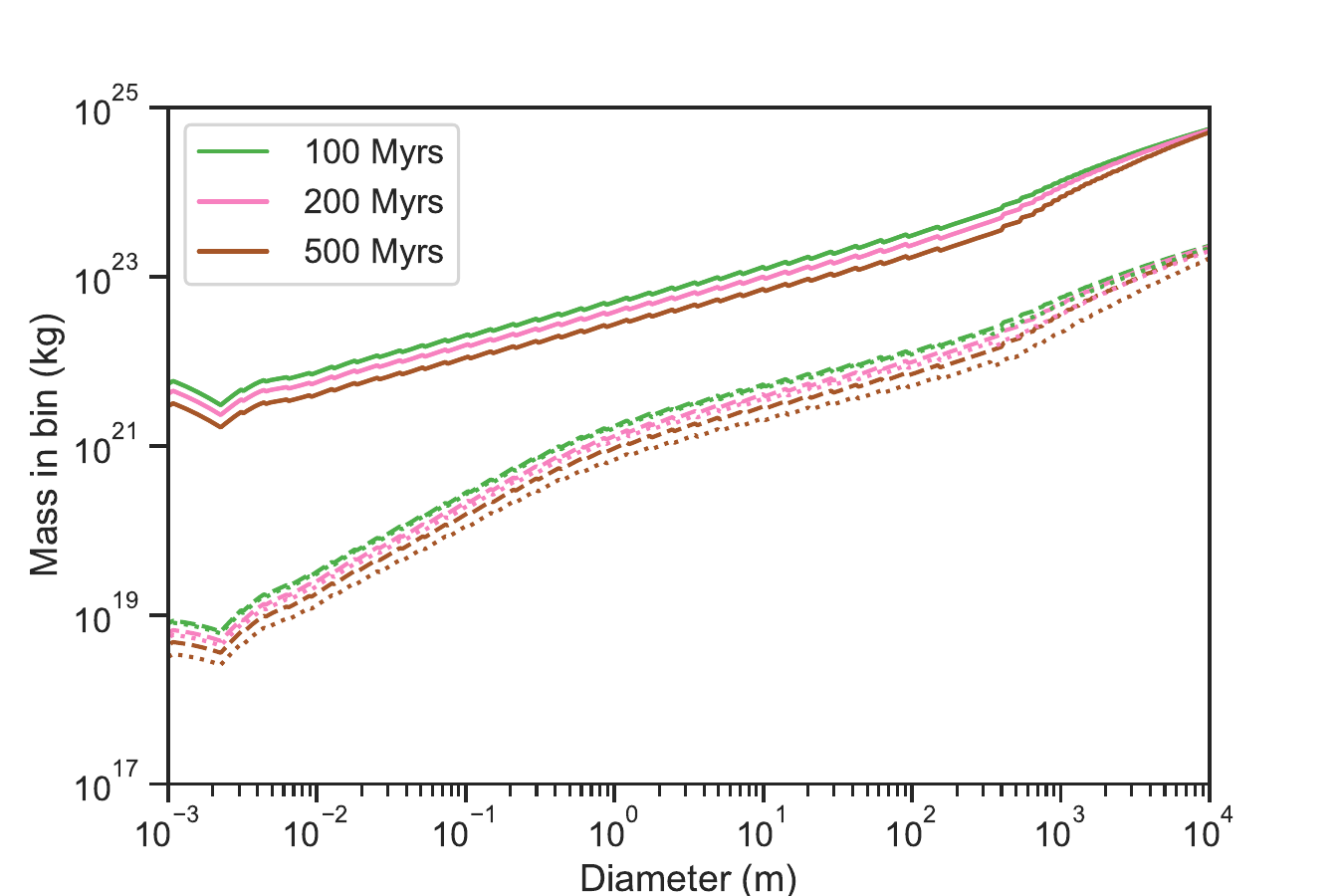}
\includegraphics[width=0.48\textwidth]{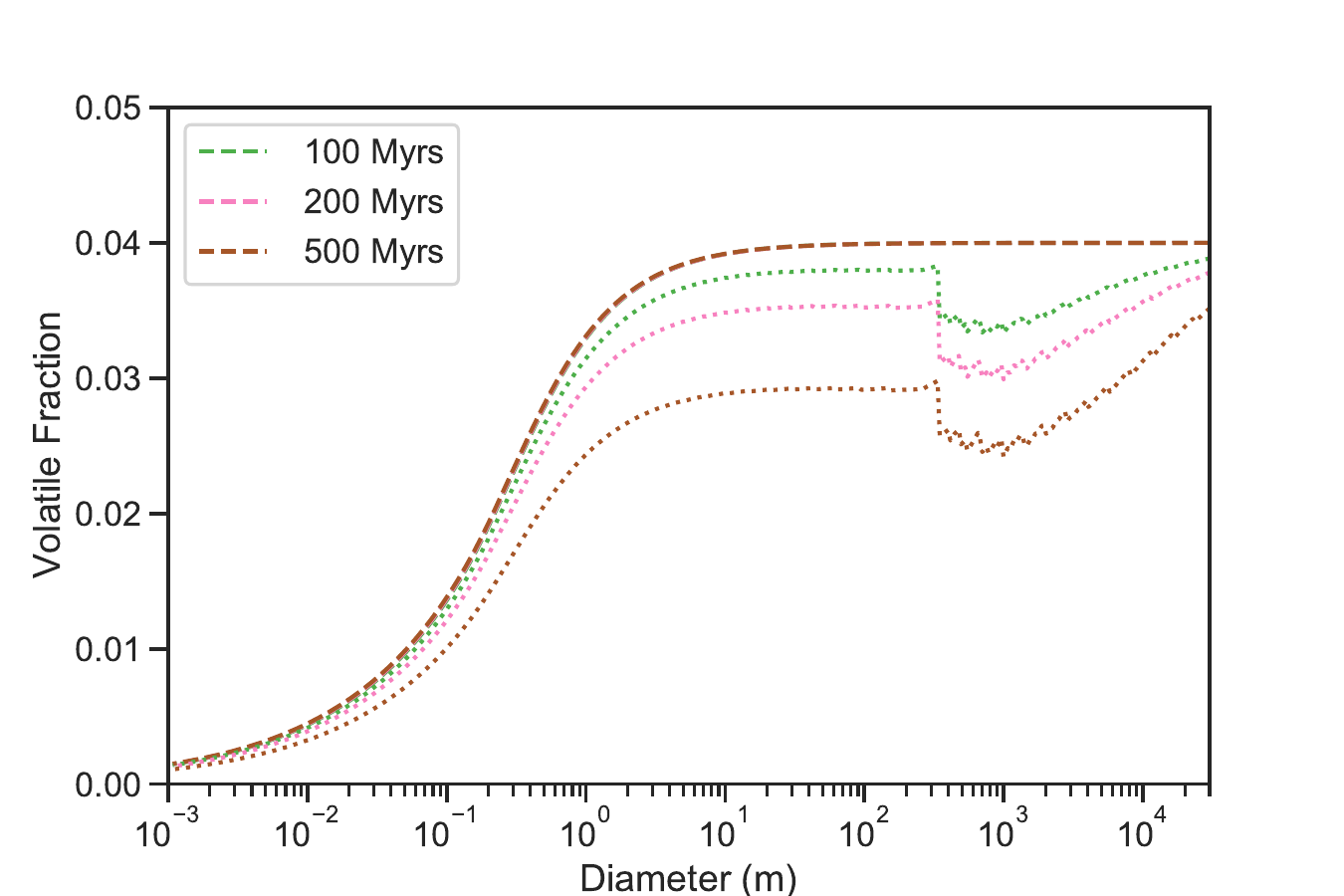}
\caption{
A comparison of the size distribution (top panel) and volatile content (bottom panel) between a simulation with both catastrophic and resurfacing collisions (dotted) or with only catastrophic collision (dashed) collisions (same simulations as Fig.~\ref{fig:example}) after 100 (green) and 200Myrs (pink). The evolution of solids, which is the same in both simulations, is shown by the solid lines in the top panel. The resurfacing collisions deplete gas in the largest planetesimals, which means that less volatiles are fed down the collisional cascade to small bodies.  }
\label{fig:example_cat}
\end{figure}


\label{sec:results}
\section{Results of the simulations for the collisional evolution of solids and volatiles in planetesimal belts}
\label{sec:results_collevol}
 The evolution of the mass in solids and volatiles  in the fiducial simulation around a solar mass star with a belt between 75 and 125au, containing $100M_\oplus$ of planetesimals in sizes between 1mm and 30km, a volatile assumed to be CO fraction of 4\% and volatiles lost from a depth $h=10cm$ following each collision, is shown in the top panel of Fig.~\ref{fig:example}. The belt location is chosen to beyond both the water-ice and CO snow-line for a solar luminosity star, such that thermal heating from the star can be neglected.  Both the solids and volatiles start with an initial $\alpha=11/6$ profile (see \S\ref{sec:belt}), but this is quickly lost as collisions erode the smallest bodies in the belt. 
 In the size domain where collisional steady-state has been reached, the size distribution slightly departs from the initial 11/6 value. This is an expected result that follows from the size-dependence of $Q_D^*$. We find a steady state slope of $\sim 1.88$ in the strength-dominated domain and $\sim 1.78$ in the gravity regime, which agrees with the theoretical prediction of \citet{wyatt11}.
 The kink or wave in the size distribution at a few mms results from the absence of grains smaller than $D_{\rm min} = 1$mm in the model considered, which leads to under and over densities of particles, as discussed in more detail in, e.g., \citet{lohne}. The artificially high minimum grain size of 1mm chosen to speed up the calculations may cause this wave to occur at larger diameters than is realistic. The very largest bodies have not yet reached collisional equilibrium, and thus, are not yet suffering catastrophic collisions and losing mass in solids. The bottom panel of Fig.~\ref{fig:example} shows the lifetime against collisions as a function of particle diameter. At 100Myrs in this disc, only bodies smaller than 1km are suffering catastrophic collisions, but this size increases with time, as larger and larger bodies suffer catastrophic collisions.

Resurfacing collisions deplete volatiles from large ($D\gtrsim 200$m) planetesimals, such that their volatile content decreases with time. This occurs on a timescale related to the resurfacing collision timescale, which the bottom panel of Fig.~\ref{fig:example} shows can be several orders of magnitude shorter than the catastrophic collision timescale, particularly for tens of kilometer sized planetesimals. This depletion leads to a sharp turn-over in the size distribution for volatiles (top panel of Fig.~\ref{fig:example} and Fig.~\ref{fig:example_cat}) at D = 200m which occurs due to the onset of resurfacing collisions. The sharp nature of this transition may not be realistic, but is an unavoidable consequence of the finite bin size used and the assumed sharp transition between collisions that do not realise any volatiles and fully shattering collisions. In reality, not only would there be a range of collision velocities, rather than the single value used here, but the transition from non-shattering to shattering collisions would occur gradually, encompassing a significant domain where collisions result in cratering rather than the assumed sharp transition at $irk$ used in these models (see also discussion in Section 5.1).

\subsection{The release of volatiles from collisions}
\label{sec:release_volatiles_collisions}

\begin{figure}
  \includegraphics[width=0.48\textwidth]{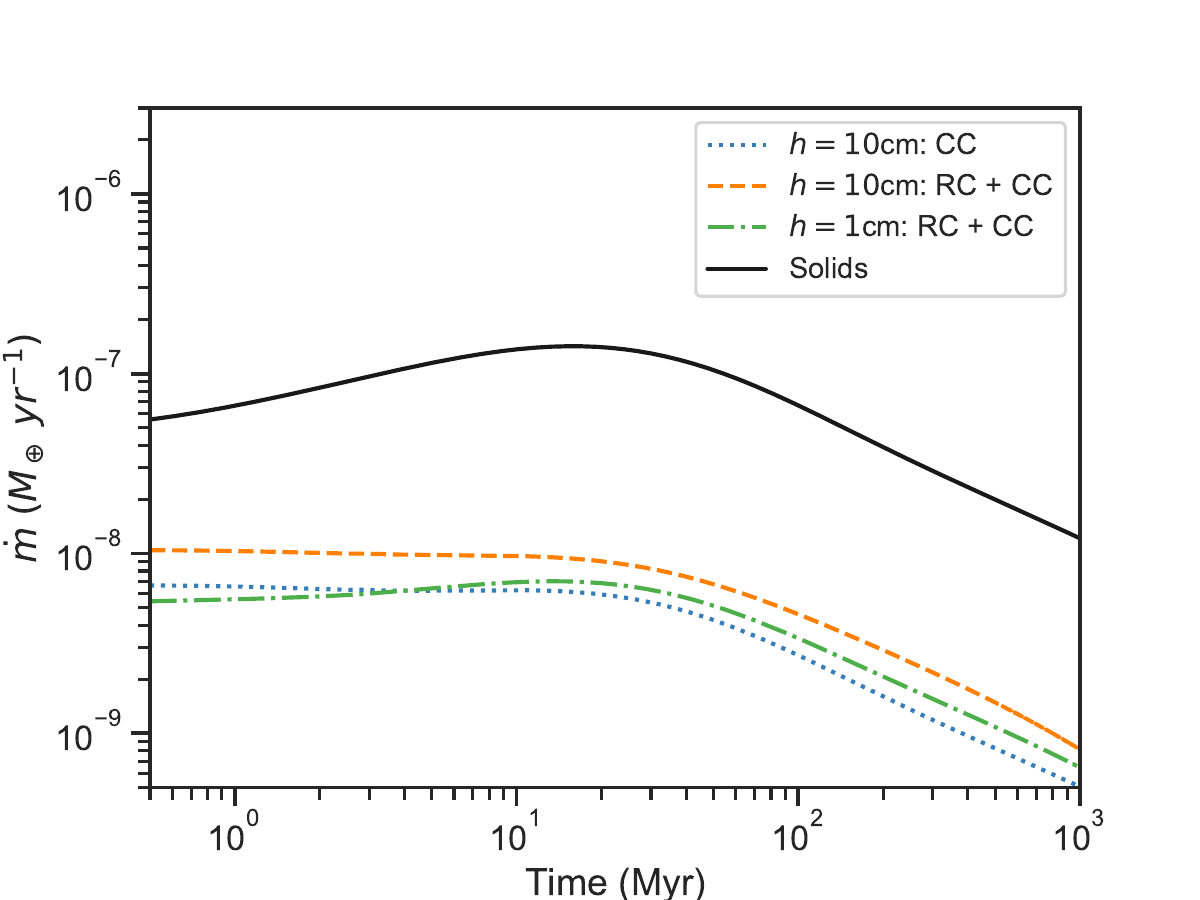}
  \includegraphics[width=0.48\textwidth]{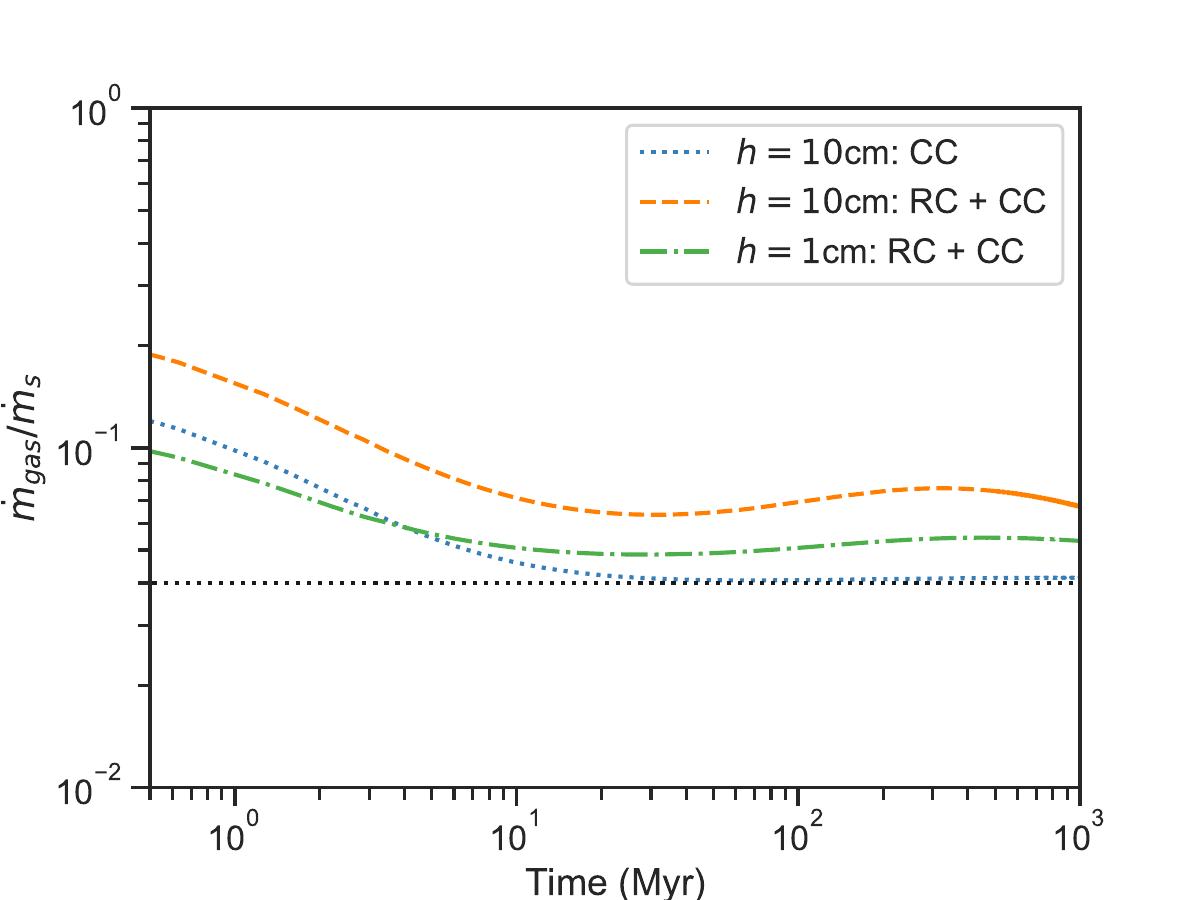} 
\caption{The gas and dust production rate as a function of time (upper panel) and the ratio of the gas production rate to dust production rate as a function of time (lower panel), for the fiducial simulation with $D_{\rm max}=30$km and $h=10$cm, with both resurfacing (RC) and catastrophic (CC) collisions  (orange dashed line) and with only catastrophic collisions (blue dotted line). The green dot-dashed line shows both resurfacing (RC) and catastrophic collisions (CC) with $h=1$cm. The dust production rate (solid line top panel),  where dust is here considered to be all objects smaller than 1mm in diameter, is the same for all simulations. The dotted horizonal line in the bottom panel indicates the initial volatile fraction of 4\%. Gas is released faster when resurfacing collisions are included, such that catastrophic collisions dominate the release of gas at late times when large planetesimals are volatile depleted (further discussion in \S\ref{sec:release_volatiles_collisions}).}
\label{fig:mgas_cat_resurf}
\end{figure}

Key for comparison with observations of gas in debris disc systems is the gas production rate due to collisions. This is traced in the numerical simulations using Eq.~\ref{eq:mgas}. The top panel of Fig.~\ref{fig:mgas_cat_resurf} plots the rate at which volatiles are released (${\dot m_{\rm gas}}$) as a function of time for a belt with the fiducial properties and $h=10$cm with only catastrophic collisions (blue dotted) compared to catastrophic and  resurfacing collisions (orange dashed). The release of gas from catastrophic collisions follows that of dust from solids (black solid line). Catastrophic collisions release volatiles at a lower rate than resurfacing collisions, whilst the rate at which gas is released by resurfacing collisions depends on their efficiency (compare $h=1$cm to $h=10$cm).

The bottom panel of Fig.~\ref{fig:mgas_cat_resurf} shows the ratio of gas to dust production. At early times ($<10$Myr), volatiles are lost from the smallest bodies ($D<30$m), as shown in the middle panel of Fig.~\ref{fig:example_cat} and, therefore, extra gas is released from volatiles relative to solids, as shown by the dotted black line in the bottom panel of Fig.~\ref{fig:mgas_cat_resurf}. This evolution could be seen as part of initialising the simulation. 

\begin{figure}

  \includegraphics[width=0.48\textwidth]{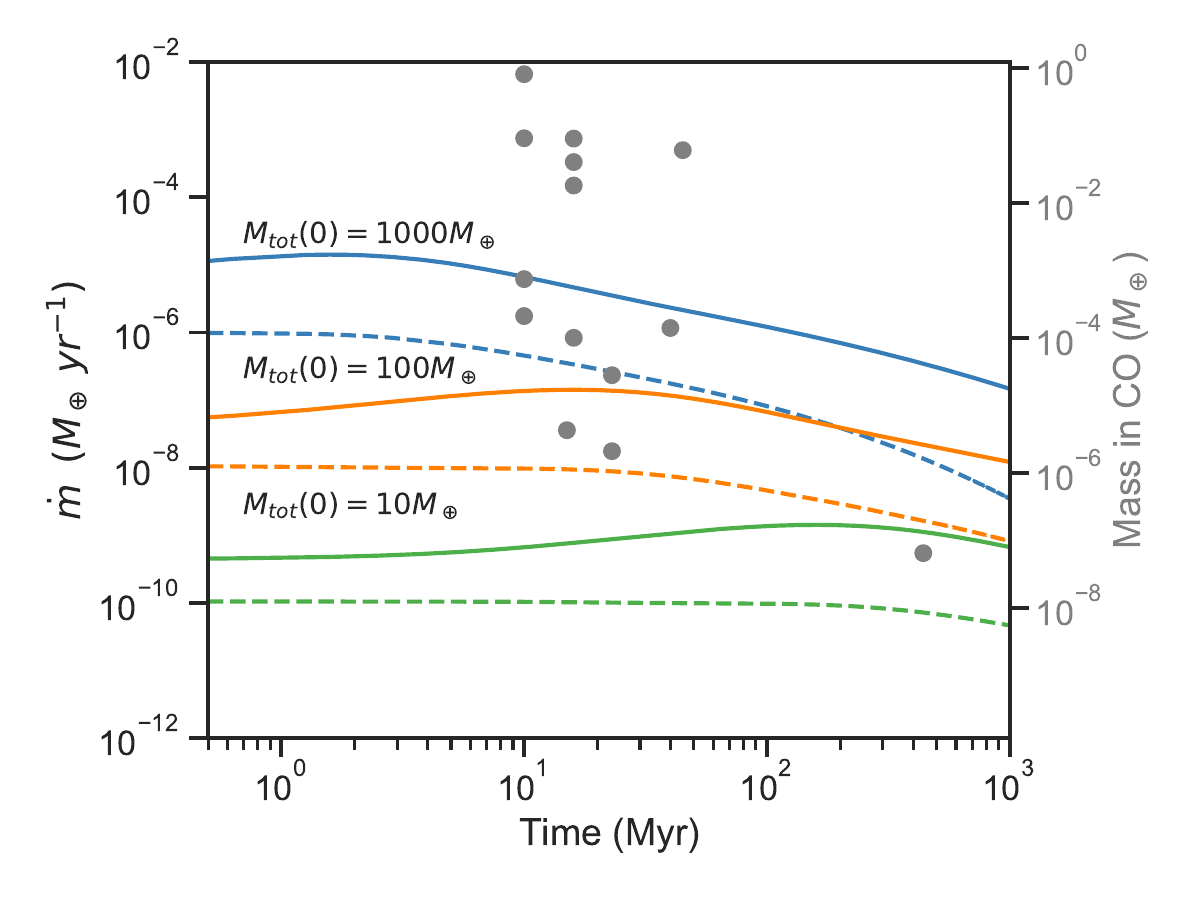}
  \includegraphics[width=0.48\textwidth]{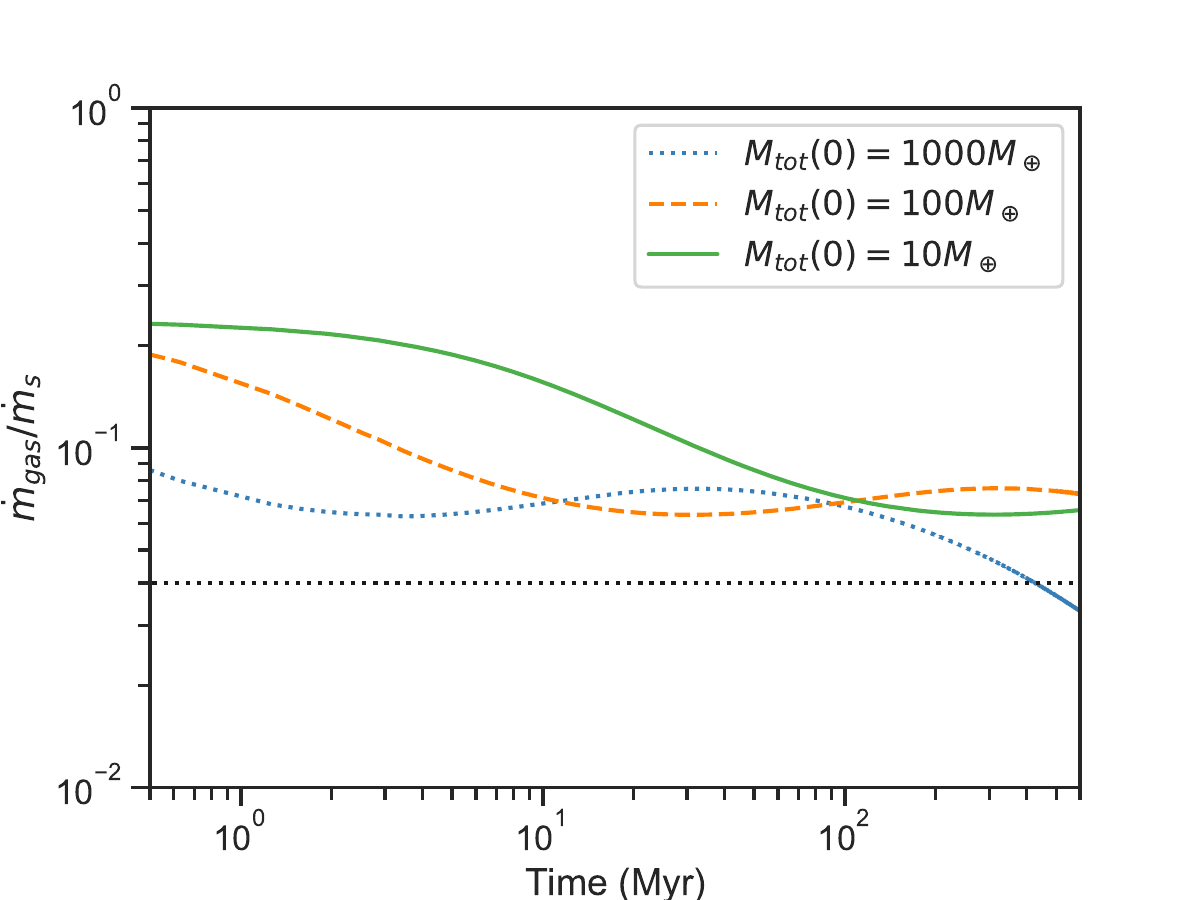}

\caption{The gas (dashed) and dust (solid) production rate as a function of time (upper panel) and the ratio of the gas production rate to dust production rate as a function of time (lower panel) for the fiducial simulation, varying the total initial disc mass ($m_{\rm s, tot}(0)=10, 100,1000 M_\oplus$.) The dotted horizontal line on the bottom plot indicates the initial volatile fraction of 4\%. The left-hand axis of the top plot additionally shows the mass in CO detected as a function of age for a selection of debris systems with CO detections (see Table~\ref{tab:CO}), noting that the ages have not been scaled for the finite proto-planetary disc lifetime. The CO mass is assumed to be depleted on a timescale of $\sim 120$yrs, indicating the many systems where shielding must occur \citep{Marino2020}. } 
\label{fig:gas_mbelt}
\end{figure}

At later times (tens of Myrs), catastrophic collisions release gas at the same fraction of the rate at which dust is released and therefore, $\frac{\dot m_{\rm gas}}{\dot m_s}$ tends to a constant value, in this case just below the initial volatile fraction of the planetesimals. When resurfacing collisions are included, however, the initial release of gas can be larger than the volatile fraction multiplied by the rate at which dust is produced, such that the fraction of gas to dust release remains above 4\% or $f_v(0)$. If the collision rate is so high that resurfacing collisions deplete the largest planetesimals of volatiles at late time, as in the simulations with $m_{s, tot}(0)=1,000M_\oplus$ the gas to dust ratio can fall below the initial volatile fraction (see Fig.~\ref{fig:gas_mbelt}).

The rate at which dust (and also volatiles) are released depends on the properties of the planetesimal belt, most notably the total initial mass, $m_{\rm s, tot}(0)$ and the size of the largest planetesimals, $D_{\rm max}$, as these influence the catastrophic collision timescale of the largest bodies present (Eq.~\ref{eq:rate_c}). Both gas and dust are released at a higher rate, for example, in more massive belts, as shown in Fig.~\ref{fig:gas_mbelt}, which shows the release of gas and dust in the fiducial simulation, compared to an order of magnitude higher and lower total initial planetesimal belt mass (noting that on sufficiently long timescales the dust evolution tends to the same level, independent of initial belt mass - see \cite{Wyatt07hot} for details).  For comparison the top plot of Fig.~\ref{fig:gas_mbelt} indicates the mass in CO detected in a selection of bright debris discs as a function of the system age, noting that the scaling of the axes is arbitrary. Details of the sample can be found in Table~\ref{tab:CO}. The intention of this plot is to indicate that most systems with CO detection are young\footnote{Most surveys of CO gas have targeted primarily young systems with the most massive debris discs, \citep[\eg][]{Moor2017, Lieman-Sifry2016, Kral2020}}, which is also when collisions release gas at the highest rate.

\section{Volatiles release due to heating from long-lived radioactive nuclides }
\label{sec:toy_cometary}
Whilst comets are in general formed in the cool outer regions of the disc, their interior temperatures can increase due to the stellar irradiation and the decay of radioactive materials in their interior. This leads to outgassing of volatiles, notably hypervolatiles such as CO and N$_2$. CO can be directly released from the sublimation of CO ice, or from where it is potentially trapped in amorphous water ice or CO$_2$ ice \citep{Davidsson2021, Lisse2022}. Whilst heating from the star can lead to release of CO \citep{Kral2020, Davidsson2021} and the decay of short-lived radioactive nuclides are important in the Solar System \citep{Prialnik1987}, this work focuses on radiogenic heating from long-lived nuclides, such as $^{40}$K, $^{232}$Th, $^{235}$U and $^{238}$U. In particular, the location of the fiducuial belt at 100au around a solar-type star was chosen such that stellar irradiation is unlikely to be the most significant contribution to heating for large comets. 

Here, we consider a simplistic model with a population of comets (planetesimals) in the belt, as described in \S\ref{sec:belt}. This model provides an estimate for the total CO released by considering that every comet releases CO at a constant rate up until a time, $t_{\rm release}(D)$, after which there is no further CO release. In total a fraction, $f_{\rm release}(D)$, of the CO in an individual comet is released. Physically, this corresponds to the time at which the comet reaches its maximum heating due to the decay of \llrscomma before their decay limits further heating. In reality the CO release from an individual comet may be peaked at earlier times, decreasing towards a time, $t_{\rm release}$ (see discussion in \S\ref{sec:discussion_activity_collisions}). The value of $t_{\rm release} (D)$ is essentially a free parameter of the model presented, but we use representative values based on more detailed simulations for Solar System comets by \cite{Davidsson2021}, where a 203km comet releases CO for approximately 25Myr, whilst a 74km comet releases CO for 30Myrs. \cite{Davidsson2021} focuses on the release of CO from amorphous water ice due to the decay of \llrscomma rather than radiation-driven processes, although the models include  time-variant protosolar heating, long-lived radionuclide heating, radial and latitudinal solid-state and radiative heat conduction, sublimation of CO ice, release of CO during segregation of CO$_2$:CO mixtures, sublimation of CO$_2$ ice, crystallization of
amorphous water ice and release of entraped CO and CO2, radial and latitudinal diffusion of
CO and CO$_2$ vapours (including mass and heat transport), and recondensation of CO and CO$_2$ vapour when applicable.

\begin{figure}
    \centering
    \includegraphics[width=0.48\textwidth]{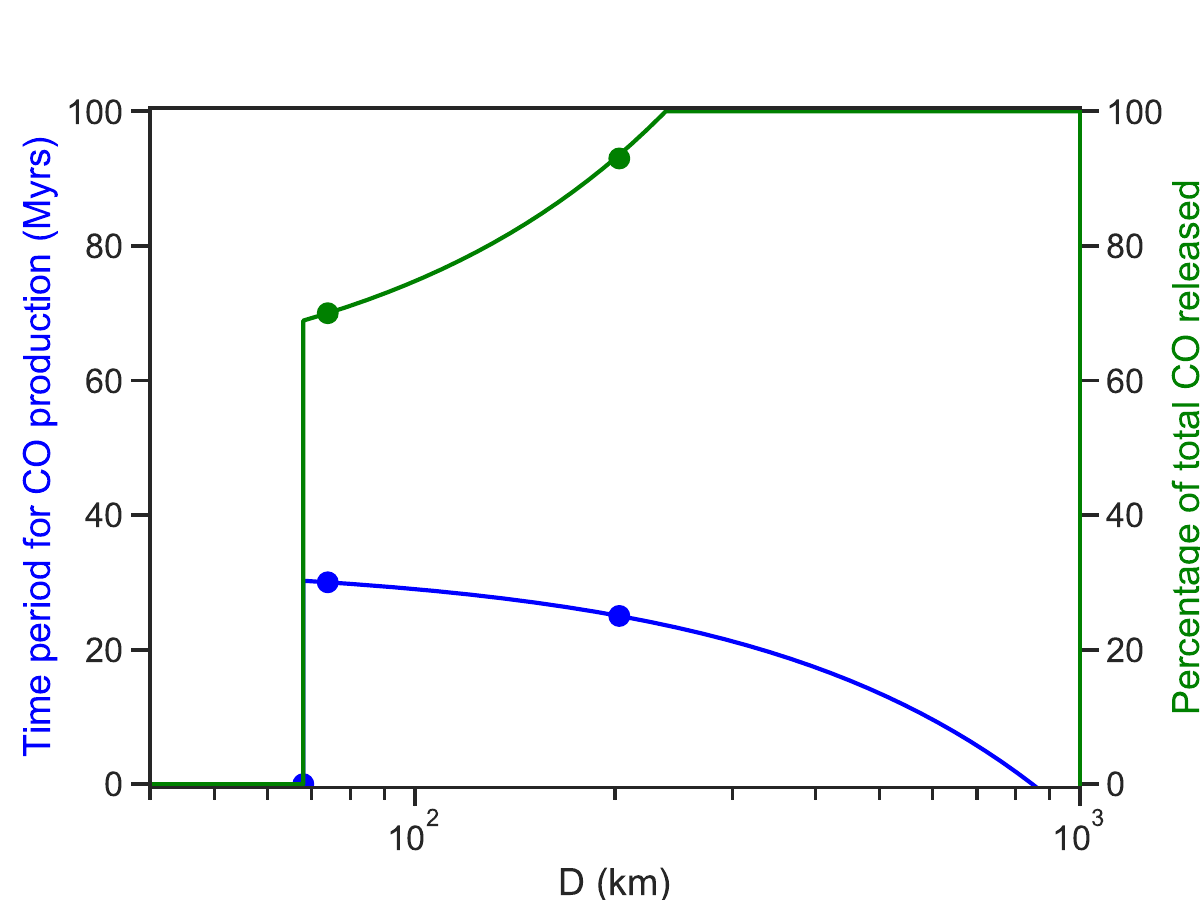}

    \caption{This figure shows the model assumptions for the time period during which CO is released, $t_{\rm release}$, as a function of planetesimal diameter, due to thermal evolution (decay of long-lived radioactive nuclides), as well as the fraction, $f_{\rm release}$, plotted on the y-axis, of the total CO present released. Further details \S\ref{sec:toy_cometary}.}
    \label{fig:early_mdot}
\end{figure}

The models of \cite{Davidsson2021} trace the thermophysical evolution of comets with a CO mass fraction of 4\%, a dust:ice mass ratio of 4 and an initial CO:H$_2$O molar ratio of 0.155, assuming Solar System abundances of long-lived radioactive nuclides and the stellar irradiation received at 23 au from the Sun. Although these models are performed closer to the Sun at 23au than our fiducial simulations at 100au, the dominant heat is the decay of \llrsfullstop Full details can be found in \cite{Davidsson2021}, including the thermal properties used for the comets. This model finds that CO trapped in amorphous water ice is not released from comets smaller than 68km due to radiogenic heating, whilst bodies larger than this limit cannot escape crystallization due to heating from \llrscomma  which leads to the release of CO to gas. This is a sharp transition in the model, as the budget of long-lived radioactive nuclides increases above a limit required to provide sufficient heating for crystallization of water ice. 
Thus, assuming the time at which all CO is released increases linearly with size $D$:    
\begin{equation}
     t_{\rm release}(D)= K_0 + K_1\,D,
     \label{eq:time_release}
     \end{equation}
where $K_0=3.8 \times 10^7$ yr and $K_1=-39 $yr m$^{-1}$. In the models of \cite{Davidsson2021}, the 203km comet releases 93\% of its total CO, whilst the 74km comet stops thermally evolving with 30\% of its total CO still present. The fraction of the total CO released is estimated assuming a linear function: 
\begin{equation}
    f_{\rm release} (D) = C_0 + C_1\,D
    \label{eq:frac_release}
\end{equation}
where $C_0 = 53$ and $C_1= 2.3 \times 10^{-4}$m$^{-1}$. Both the time and the fraction of CO released are shown as a function of diameter in Fig.~\ref{fig:early_mdot}. No dependence on the distance to the star is assumed in this simple model. We acknowledge that the exact timescales on which CO release continues, the proportion of the total CO released and the minimum size for which long-lived radioactive nuclides can heat sufficiently that any CO is released would all vary with many of the free parameters used in the \citep{Davidsson2021} models, see \S\ref{sec:discussion_validity}.

In order to implement the release of volatiles due to thermophysical evolution, each mass bin is considered to release volatiles at a constant rate up until a time $t_{\rm release}^k$. The rate at which mass in the $k$-th bin releases volatiles is given by the total mass released divided by the time period over which it is released, 
\begin {equation}
\dot{m}_{k, \rm a} = \frac {f_{{\rm v}, k} K M_{\rm s, k}^{2-\alpha} \delta f_{\rm release}^k}{ (1-f_{{\rm v}, k}) t_{\rm release}^k}, 
\end{equation} 
where $K$ is the constant defined in Eq.~\ref{eq:m_s}. We note that this equation is based on the premise that there is no evolution of the mass in the belt, such that the power-law for the size distribution with constant $\alpha$ continues to apply. The total volatile release is the sum over all bins, noting that volatiles are only released up to a time, $t_{\rm release}^k$, and that no volatiles are released for bodies smaller than $D<68$km, as these bodies do not have a sufficient budget of \llrs to lead to mobilisation of CO:

\begin{equation}
   \dot{m_{\rm a}} = \Sigma_{k=1}^{i_{\rm 68}} Z_k(t) \,\dot{m}_{k, \rm a} ,
    \label{eq:mdotearly}
\end{equation}
where $i_{\rm 68}$ labels the smallest mass bin to release CO, in this case $D=68$km, $Z_k$ is a function which equals 1 for times, $t$, shorter than $t_{\rm release}^k$ and otherwise zero. 

 We acknowledge that this basic model falls short of more detailed thermal evolution models, but it is intended to provide an indication of the probable levels of gas release, which can be compared with the release from collisions.

\begin{figure}

  \includegraphics[width=0.48\textwidth]{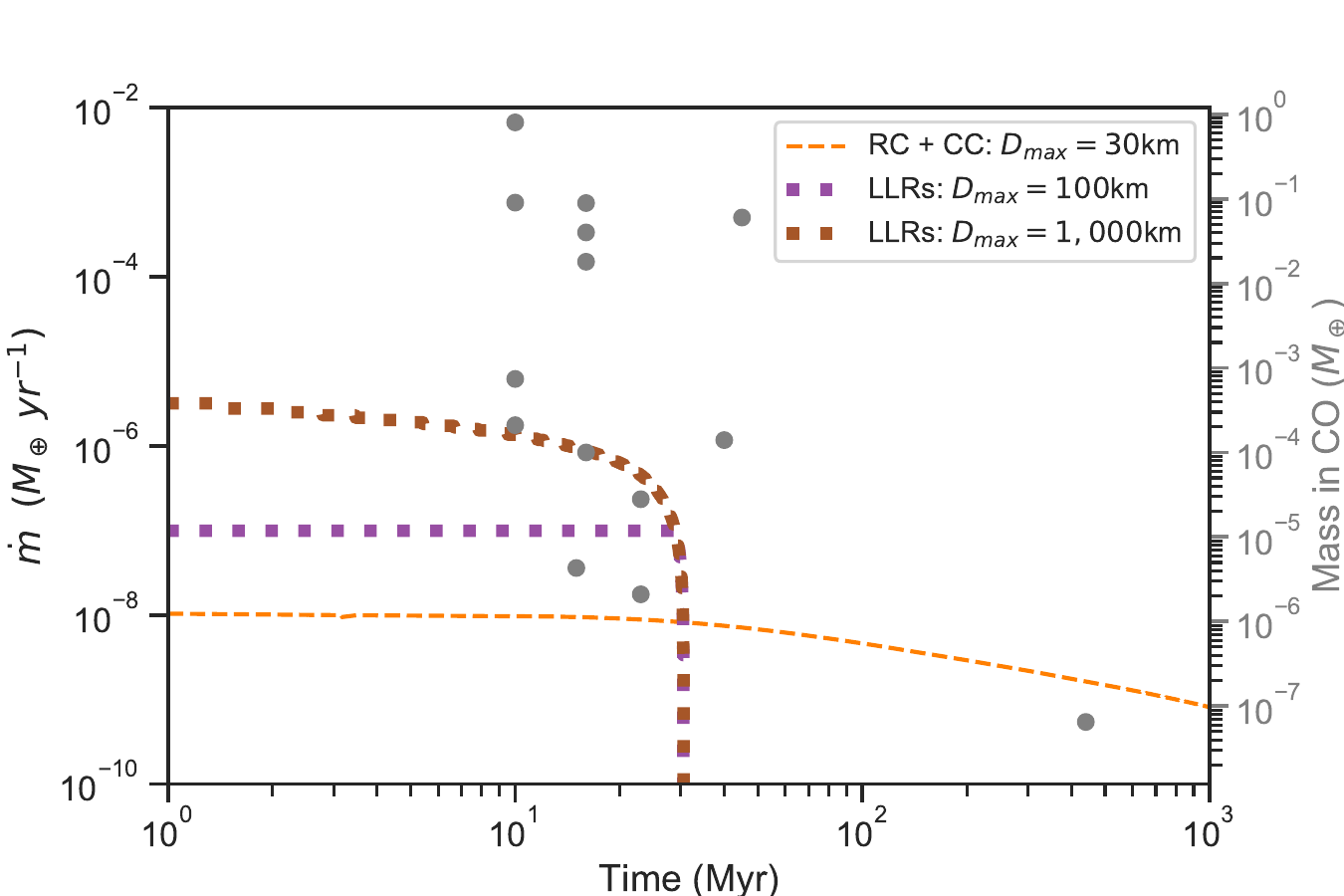}

\caption{The release of gas from resurfacing and catastrophic collisions (dashed lines) and radiogenic heating (dotted lines) as a function of time, for the fiducial planetesimal belt (see \S\ref{sec:numerical} for full parameters) centred at 100au, varying the size of the largest planetesimal, $D_{\rm max}$ from 30km (no gas production due to radiogenic heating: orange) to 100km (purple) and 1,000km (brown). The total initial belt mass, $m_{\rm s, tot}(0)$, is scaled with $D_{\rm max}$ to retain the same initial dust production rate, indicating a potential test for the size of the largest planetesimal, based on different predicted gas production rates. The right-hand axis additionally shows the mass in CO detected as a function of age for a selection of debris systems with CO detections (see Table~\ref{tab:CO}, as in Fig.~\ref{fig:gas_mbelt}.) }
\label{fig:gas_cometary}
\end{figure}

\subsection{A comparison between the volatile release from thermophysical evolution (\llrsbracket and collisions. }
\label{sec:results_early}
The model described in \S\ref{sec:toy_cometary} is used to quantify the gas released from a planetesimal belt in which all (many) of the comets are active, with the focus being the late time thermal evolution due to the decay of long-lived radioactive nuclides, leading to the release of CO trapped in CO$_2$ or amorphous water ice.  Fig.~\ref{fig:gas_cometary} shows the total CO production rate from the fiducial belt (see \S\ref{sec:numerical}), assuming an average rate.  This release in gas depends crucially on the total mass in large planetesimals. It is only planetesimals larger than $D=68$km that contain a sufficient budget of long-lived radioactive nuclides in this model to contribute to the release of CO. This limit depends on the efficiency of heating or cooling and the total budget of long-lived radioactive nuclides, but is unlikely to change by orders of magnitude for Solar System budgets of \llrsfullstop The CO in the 68km planetesimals is released at the latest times, which in this model means that no CO is released due to thermal evolution after around 30 Myrs, noting that in reality this is unlikely to be a hard cut-off and the exact time at which it occurs may not be well predicted by this model (see further discussion in \S\ref{sec:discussion}).

Crucially in this simple model the gas released depends on the population of large planetesimls. 
Fig.~\ref{fig:gas_cometary} shows that the gas production rate depends on the mass in large planetesimals present. When the largest planetesimal is increased from 100km (purple dotted line) to 1,000km (brown dotted line), whilst maintaining the same mass in dust, the gas production rate increases by over an order of magnitude. This is plausibly an important test of whether large planetesimals are present in the planetesimal belt.

Both radiogenic heating and collisions release volatiles at rates which are within the same order of magnitude (see Fig.~\ref{fig:gas_cometary}). The gas production rates are broadly dominated by the availability of CO within the planetesimals in the belt within this simple model.

\section{Discussion}
\label{sec:discussion}
This work presents a simple model that aims to quantify the production of gas (most notably CO) from planetesimal belts, based on the parameters of the planetesimal belt. CO is released at early times due to thermal evolution powered by the decay of long-lived radioactive nuclides. This is compared to the release of CO due to both resurfacing and catastrophic collisions, following the collisional evolution of the planetesimal belt. The models point to the importance of thermal evolution in young planetary systems ($<30$Myr), whilst collisional gas production can be maintained on Gyr timescales.

In the following sections, we first discuss the validity of the model presented, highlighting the many simplifying assumptions. We then discuss whether the model presented here can be used to distinguish the dominant method of gas production in debris systems and whether that is related to radiogenic heating, resurfacing or catastrophic collisions. 

\subsection{Validity of Model}

\label{sec:discussion_validity}
The biggest simplification of the model presented here is
the lack of any attempt to model the interior structure of
the cometary bodies. This is crucial for the comet’s thermal
evolution, the location of gases within the comet, the struc-
ture of the comet and thus, its ability to release gas during
collisions. Whilst there have been significant advancements
in our understanding of cometary interiors in recent years
\citep[\eg][]{Steckloff2021, Malamud2022, Davidsson2021}, several key open questions remain and simulations
are computationally expensive even for a single comet. The
aim here is to consider the population of planetesimals as
a whole. Key changes that would influence these models include the thermal conductivity of the comets. The model
presented here is based upon the timescales found in the simulations of \citep{Davidsson2021} who adjust the thermal conductivity as tabulated for H5 ordinary chondrite, amorphous,
cubic and hexagonal water ice \citep{YomogidaMatsui1983, Klinger1980, Kuhrt1984}, CO and CO$_2$ \citep{Giauque1937} for the anticipated porosity of comets. This matches
timescales predicted for the removal of hypervolatiles from
Arrokoth \citep{Steckloff2021}. A reduced thermal conductivity would minimise the energy re-radiated in the infrared,
such that heating is faster and CO loss occurs earlier. This
would make it harder for CO loss to be sustained on long
timescales, as required to explain gas detections in older exoplanetary systems. We acknowledge the importance of
these poorly known parameters and the exact details of the
thermal evolution model in determining the release of CO.

A second important limitation of the model regards the interior structure of the comet, in particular, the location of hypervolatiles following initial thermal evolution. A typical outcome of models with radiogenic heating is that activity removes volatiles from the core, leaving a cold volatile-bearing mantle intact \citep{Davidsson2021, Malamud2022}. Collisions can then release CO from this cold outer crust whenever they occur. 
 
In particular this would influence the simple model for the release of gas from collisions, presented in \S\ref{sec:release_CO}. This model is clearly an over simplification of reality, retaining only the baseline assumption that the release of gas is proportional to the surface area of the fragments. If different collision strengths or velocities are more/less efficient at releasing volatiles, this would significantly influence the overall gas production rate predicted by the models. 

The interior structure of the comets has a crucial influence on the ability of collisions to break them apart. This is modelled here by the simplistic prescription for $Q_D^*$ based on SPH modelling of collisions of icy and rocky bodies from \cite{benzaphaug}, however, this may take a different form if comets are formed predominantly via pebble accretion or are highly porous \citep{Blum2017, Jutzi2017, Jutzi2017a, Davison2010, Krivov2018}. The timescales involved in the collision models presented here depend crucially on the strength of the planetesimals, thus, these timescales could potentially increase (decrease) significantly in an improved collision model. However, in the collision models, these timescales would scale in the same manner for both gas and dust production, such that even in this case, the model presented here has the power to distinguish whether the gas production is purely related to catastrophic collisions.

The efficiency at which collisions release volatiles is parameterised in a very simple manner in the model presented here (see \S\ref{sec:release_Cat} and \S~\ref{sec:release_resurf}). It is based upon the premise that the fractional release of volatiles ($\chi_k$) depends on the surface area of the planetesimal fragments. Whilst this appears a reasonable broad assumption as most thermal processes depend on the surface area available for heating or cooling, it is clear that the story is more complex for collisions which occur, for example, only in a particular region of the planetesimal or in a non-axisymmetric manner. The mode of heat transport through the body during such collisions is unclear. In this work this lack of knowledge is parameterised by the free parameter $h$. However, we acknowledge that this free parameter may have values which are orders of magnitude different depending on whether the key process is thermal heating from stellar irradiation, UV desorption or the heat released during collisions.

The situation for resurfacing collisions is even less clear. The model presented here assumes a size distribution of fragments which each lose volatiles before the body reaccumulates, with the absence of knowledge of the timescale for this re-accumulation parameterised in the free parameter $h$. Again, both the timescale, the size distribution and the symmetric nature of this process are poorly understood. 
 
We also note that in a realistic system it is unlikely that volatile release is equally efficient from the fragments that reaccumulate to form large bodies in the gravity regime in both resurfacing and catastrophic collisions and from the fragments produced in catastrophic collisions, or in other words $h$ is unlikely to take the same value for both collision types, as assumed here. Accounting for this potential additional volatile loss would only have a relatively minor effect, leading to earlier gas release from the belt.

The collision models presented here ignore the possibility of cratering collisions, which is clearly an oversimplification. Detailed numerical investigations have indeed shown that cratering impacts can play a significant role in the collisional evolution of debris discs, leaving an imprint in the system's global particle size distribution \citep{Thebault07}.
In the present case, we expect such collisions to increase the release of volatiles to gas by chipping away at the outer layers of planetesimals. This means that the presented results do probably underestimate the level of gas production by collisions at early times, while probably overestimating the duration of the gas production phase (because the reservoir of volatiles would be drained more quickly). The inclusion of cratering impacts would probably also smooth out any unphysical transitions in the size distribution (as seen for example in Fig.~\ref{fig:example}). In the present model there is indeed, for a given collisional target, an abrupt jump around the projectile mass $M_rk$, between $M>M_{rk}$ projectiles that can fully shatter the target into a distribution of fragments that will all release gas and $M<M_{rk}$ projectiles that have zero effect on the target.
An additional crucial simplification of the models is the assumption that the largest post-impact fragment is always half the mass of the target, whereas in reality this size decreases strongly depending on the energy imparted \citep[\eg][]{Leinhardt09}.


The models for the thermal evolution of comets are based on a Solar System-like budget of long-lived radioactive nuclides. It is thus, crucial to question whether Solar System-like budgets are likely to be typical across exoplanetary systems. Whilst there is a base-line contribution to long-lived radioactive nuclides from the nuclear supply of the galaxy, their budget can be enriched by supernovae, either directly, or by enrichment of the star-forming molecular cloud. These processes are ubiquitous and all exoplanetary systems will be enriched to a certain degree in long-lived radioactive nuclides. Abundances of Thorium in sun-like stars suggest that most exoplanetary systems around sun-like star have similar, if not higher abundances of \llrs \citep{Unterborn2015, Botelho2019}. The story for the more volatile $^{40}K$ may, however, differ, with galactic chemical evolution models  suggesting that Solar System-levels of $^{40}K$ occur in about 1/80 exoplanetary systems \citep{FatuzzoAdams2015}. This is interesting to note, as whilst a reduced budget of long-lived radioactive nuclides would increase the minimum size heated sufficiently to lead to CO out-gassing, it is plausible that larger comets could continue to release CO on longer timescales than in the models presented here.

The models presented here ignore the presence of short-lived radioactive nuclides, such as $^{26}$Al, as the budget of these in exoplanetary systems is unknown, which some studies suggesting that Solar System-like budgets are typical, whilst others that very few systems are enriched at similar levels as the Solar System \citep[\eg][]{Gounelle2015, Lichtenberg2016, Kuffmeier2016, Young2014}. The decay of short-lived radioactive nuclides was potentially important for Solar System comets \citep[\eg][]{Parhi2023, Mousis2017}, although the presence of amorphous water ice could suggest a limited budget of $^{26}$Al at formation \citep{Prialnik1987}.

The model presented here essentially ignores any thermophysical evolution prior to the end of the gas disc lifetime and assumes that the planetesimals are fully formed at this point. We acknowledge here that planetesimal belts may not be collisionally active (\ie {\it stirred}) at the end of the gas disc lifetime, with rather a continued period of growth prior to self-stirring, as discussed in detail in \citep[\eg][]{grantstirring}.

The model presented here treats the thermophysical evolution and collisions as separate processes. In reality both processes may act on the same bodies. In which case, the late time collisional gas production may be significantly depleted due to the fact that volatiles have been lost from the largest planetesimals due to thermal evolution. The retention of some volatiles in 68-100km planetesimals (see Fig.~\ref{fig:early_mdot}) and all volatiles in smaller planetesimals allows for the continued gas production from collisions on long timesacles, as relevant for example for Fomalhaut.

\subsection{Radiogenic heating or collisions?}
\label{sec:discussion_activity_collisions}
This paper highlights three main channels for the secondary production of gas in debris discs; radiogenic heating, catastrophic or gentler resurfacing collisions. The model presented here for the thermophysical evolution focuses on the heating due to the decay of long-lived radioactive nuclides, whilst we acknowledge here that for comets sufficiently irradiated by their host stars, external heating may also play a role. All three processes are able to sustain the release of CO at levels similar to those required to explain the observations, around $10^{-6}-10^{-4}M_\oplus$ dissociating in $\sim 100$yrs\footnote{In belts with higher CO masses, CO likely has a longer photodissociation timescale as it is self-shielded or shielded by other species \citep{Marino2020}, therefore it is difficult to determine the CO release rate in those systems or assess whether it is of secondary origin.}, i.e $10^{-8}-10^{-6}M_\oplus$yr$^{-1}$ (see Fig.~\ref{fig:gas_mbelt}), depending on the properties of the planetesimal belt. The key difference between the decay of \llrs and collisions here are the timescales, with radiogenic heating only leading to gas production at early times. Thus, whilst the decay of \llrs can explain the detection of gas in the majority of debris systems which are around young (<~30Myr) stars, the detection of CO gas in older planetary systems, provides a key test. These include systems such as 49 Ceti at 40Myr or HD 21997 at 45Myr, where there is sufficient uncertainty in whether shielded CO could have survived or radiogenic heating could continue on slightly longer timescales than those used in the model presented here. However, we have to acknowledge uncertainties in the model presented here. Significantly older systems such as Fomalhaut at 440Myr \citep{Mamajek2012}, on the other hand, render the current release of gas from the decay of long-lived radioactive nuclides or the survival of CO from an earlier epoch unlikely. 

In the model used here the maximum timescale for CO release from radiogenic heating depends on when the maximum temperature of the planetesimals is reached, which in turn depends on the exact structure and cooling of the cometary bodies. In the models of \cite{Davidsson2021} a decrease in the dust:water-ice ratio, accompanied by an increase in the CO:H$_2$O would reduce the rate of heating from LLRs and increase the timescale for which CO could be released (by increasing the total amount of CO to be released). Whether this could be increased sufficiently to produce gas at the rate required for Fomalhaut after 440Myr is not clear. 

An alternate explanation for the gas production in the Fomalhaut system is the release of volatiles following heating by stellar irradiation. \cite{Davidsson2021} show that a 200km body continues to lose CO from CO ice for 200Myr when irradiated by the Sun at 23 au, in a similar manner to that mentioned in \cite{Kral2021}. 
If this CO is released, the same irradiation would occur at 93 au in the Fomalhaut system ($L_*=16.6L_\odot$), whilst at the location of the Fomalahaut belt (130-150au), the irradiation is reduced to just under half and in principle CO loss could continue for just over twice as long ($\sim$400-500Myr).  However, the rate of release of CO would potentially be lower. A simple estimate finds a constant average rate of $10^{-9}M_\oplus$ yr$^{-1}$, assuming that the total mass in the Fomalhaut system in bodies up to 300km is 63$M_\oplus$ \citep{Krivov2021}, with a CO mass fraction of 4\%. This low release rate may just be able to explain the low mass of CO ($10^{-7}M_\oplus$ \cite{Matra2017_fomb}, dissociating in $\sim 100$yr).


On the other hand, if CO remains present within the planetesimals in the Fomalhaut belt, collisions will continue to release CO for over 440Myr. Assuming an initial CO fraction of 4\%, the belt's location (143au), width (13.6au) and predicted total planetesimal mass ($1.8M_\oplus$) in bodies up to 0.3km \citep{Krivov2021}, the collisional gas production (both catastrophic and resurfacing collisions) would be $10^{-9}$ $M_\oplus$yr$^{-1}$.

Thus, whilst the decay of long-lived radioactive nuclides is unlikely to continue on sufficiently long timescales to explain the gas production observed at Fomalhaut, if the Fomalhaut planetary system had a substantially lower initial budget of long-lived radioactive nuclides (such that volatiles can survive in large planetesimals on long timescales), it remains plausible that the low (compared to other debris discs with CO detection) CO mass observed at Fomalhaut could be released due to the stellar irradiation slowly heating and mobilising the CO ice. This explanation, however, depends crucially on the presence of large (hundreds of km) planetesimals. Collisions, are able to sustain a low rate of gas production for the age of Fomalhaut without the presence of large planetesimals.

Thus, this work suggests that both thermal evolution and collisions lead to the release of gas in debris disc systems, with thermal evolution dominating at early times, but less likely to explain CO in old planetary systems such as Fomalhaut. 

\subsection{Resurfacing or Catastrophic Collisions} 
\label{sec:discussion_resurf_cat}
It is currently difficult to find observational evidence that the gas observed in debris systems is produced in resurfacing collisions, rather than catastrophic collisions. One prediction of the models presented here is that the rate of gas production from catastrophic collisions is proportional to the dust production rate (infrared emission), whilst for resurfacing collisions it depends additionally on the population of large planetesimals. At face value the observed population of debris discs with and without CO detections could be seen as evidence in support of resurfacing collisions as there does not appear to be a direct correlation between infrared emission and mass in CO detected, see \cite{Marino2020} for a detailed discussion. However, this lack of correlation can plausibly be explained by two things. Firstly, the observed CO may be shielded and thus, not proportional to the CO production rate. Secondly, the CO production rate from catastrophic collisions may be proportional to the dust production rate, but will depend additionally on other parameters which can vary between systems, such as the CO fraction of the planetesimals. Thus, it is not currently possible to use the observed population to rule out the potential importance of resurfacing collisions in CO production.

From a theoretical perspective, however, it seems likely that non-catastrophic collisions, not just resurfacing collisions, but cratering or other non-catastrophic collisions occur in planetesimal belts. 
Whilst this model does not explicitly include cratering collisions, these collisions could potentially release CO at earlier times, as the more frequent cratering collisions chip away at the outer layers of the planetesimals. However, the bulk of the CO, trapped in the deep interior would still need to wait for a shattering or resurfacing collision to be released.

\subsection{How big are the largest planetesimals in debris discs?}
\label{sec:discussion_howlarge}
This is a crucial question, as highlighted for example by \cite{Krivov2021}, which determines the long-term evolution of debris systems. Whilst the Solar System's debris belt contains large (D$\sim 1,000$kms) planetesimals such as Pluto, the presence of large ($D>100$km) planetesimals contradicts observations that indicate fewer old systems have high infrared luminosities from dusty planetesimal belts, as predicted by the collision evolution of belts containing only small planetesimals \citep{su06,wyatt07, Krivov2021}. Additionally, the mass budget in planetesimals required for the brightest observed debris systems would exceed that of the solid component of proto-planetary discs or exoplanet population. 

Gas production from radiogenic heating depends crucially on the population of large planetesimals and thus, provides, a test for the size of these bodies in debris discs. This is clearly seen in Fig.~\ref{fig:gas_cometary}, where the gas production rate is significantly higher, for the same dust production rate, when the population of larger planetesimals is increased. As resurfacing collisions also only occur for large planetesimals, if planetesimal belts do not contain a population of large ($>$ tens of km) planetesimals, catastrophic collisions are more likely to dominate the observed release of CO.

In the collisional production model, there is another important parameter controlling the system's evolution, namely the average eccentricity $e$ of planetesimal orbits. This parameter indeed determines mutual impact velocities and thus the outcome of collisions. We took $e=0.1$,the typical value usually assumed for debris-producing discs \citep{thebault2009}, but lower values have been inferred for some discs, which would lead to a less intense but longer-lasting collisional activity \citep{loehne2012}.

\subsection{The composition of comets, as derived from gas observations} 
\label{sec:discussion_composition}
The detection of individual gases released from comets in debris discs provides a unique opportunity to probe the composition of comets in exoplanetary systems, in comparison to our Solar System, as in \eg \cite{Matra2017_fomb}. This work highlights the difficulties in using observed CO as a probe of the total CO content of planetesimals. Thermal evolution is likely to have played a significant role in reducing the initial volatile content of comets, even during the primordial disc phase, for comets both in the Solar System and in exoplanetary systems \citep{Davidsson2021, LichtenbergKrijt2021}. Additionally, the models presented here, notably Figs.~\ref{fig:mgas_cat_resurf},~\ref{fig:gas_mbelt}, show that when resurfacing collisions are considered the ratio of the gas to dust production rate can be significantly above (at early times) or below (at late times) the CO content of the planetesimals.

\section{Conclusions}
\label{sec:conclusions}
The observation of gas in traditionally gas-poor debris disc systems provides crucial clues regarding the evolution of volatiles within planetary systems. Here, we compare a model that predicts the secondary release of gas from planetesimal belts due to heating from the decay of \llrs to a model for the collisional production of CO in both catastrophic and resurfacing collisions. The release of gas from catastrophic collisions follows the dust evolution of the belt, whilst non-catastrophic collisions, such as resurfacing (or shattering) collisions in large (hundreds of kms) planetesimals contribute to the early release of gas at higher rates than with only catastrophic collisions. We predict  the gas release from collisions as a function of properties of the planetesimal belt. The release of gas from resurfacing collisions depends crucially on the presence of large (hundreds of km) planetesimals and means that the observed rate of CO release compared to the dust production is not always a good probe for the CO content of the comets.

Radiogenic heating from the decay of isotopes such as $^{40}$K, $^{232}$Th, $^{235}$U and $^{238}$U can lead to the heating of comets and CO gas production rates comparable to those required to explain the observations, if planetesimal belts contain tens to hundreds of kilometer planetesimals. Radiogenic heating has the potential to explain the CO observed in all young ($<50$Myr) planetary systems, whilst the presence of CO gas in older planetary systems, most notably Fomalhaut at 440Myr \citep{Matra2017_fomb}, is readily sustained by collisions. We highlight the potential importance of the slow penetration of stellar irradiation to the deep interiors of comets, as suggested by \cite{Kral2021}, particularly for old planetary systems, such as Fomalhaut.

\section{Data Availability}
The data and codes used in this manuscript can be found at \url{https://github.com/abonsor/coll_gas}

\section{Acknowledgements}
AB acknowledges the support of a Royal Society University Research Fellowship, URF\textbackslash R1\textbackslash 211421. SM is supported by a Royal Society University Research Fellowship (URF-R1-221669). We acknowledge fruitful discussions with Uri Malamud and J\"{u}rgen Blum. Parts of the research were carried out at the Jet Propulsion Laboratory, California Institute of Technology, under a contract with the National Aeronautics and Space Administration

\begin{table*}
\caption{List of Variables }
\begin{tabular}{ |p{2cm}|p{1cm}|p{14cm}}
\hline

$\alpha$ & & The power index of the mass distribution of planetesimals in the belt $n(M)dM\propto M^{-\alpha}$ \\

$a$ & $0.3$ & The dispersal threshold contains the term $D^a$ see Eq.~\ref{eq:qdstar}\\

$b$ & $1.5$ & The dispersal threshold contains the term $D^b$ see Eq.~\ref{eq:qdstar}\\

$\chi_k(f_{{\rm v},i})$ & & The fraction of the volatile mass lost to gas following a collision, which is explicitly a function of the volatile fraction of the original fragments in the $i$-th bin \\
$C_0$  && A constant which describes the linear increase in the fraction of CO released by LLRs in Eq.~\ref{eq:frac_release} \\
$C_1$&& A constant which describes the linear increase in the fraction of CO is released by LLRs in Eq.~\ref{eq:frac_release} \\
$\delta$ & & The logarithmic spacing in mass bins used in the numerical model\\

$\delta_t$ & yrs& The timestep used in the numerical simulations \\
$D_{\rm min}$ &m&The minimum diameter present in the collisional cascade or the blow-out size \\
$D_{\rm max}$ & m & The maximum diameter present in the collisional cascade \\
$D_k$ & m& Diameter of planetesimals in the $k$-th bin \\
$D_w$ & m & The minimum in the dispersal threshold \\
$e$ && The average eccentricity of colliders\\
$f_{{\rm v},k}$ & & volatile fraction of planetesimals in the $k$-th bin \\
$f_{\rm release}$ &&The fraction of the total CO which is released due to the decay of LLRs \\
$F_s(k-i)$& & The redistribution function for solids, or the fraction of the solid mass leaving the $i$-th bin from collisions that enters the $k$-th bin \\
$F_{\rm v}(k-i)$& & The redistribution function for volatiles, or the fraction of the mass in volatiles leaving the $i$-th bin from collisions that enters the $k$-th bin \\

$h$ & m& The depth from which volatiles are released in a collision \\ 
$K$ && The constant of proportionality in the mass distribution of planetesimals in the belt Eq.~\ref{eq:m_s} \\
$K_0$  && A constant which describes the linear increase in the time up until which CO is released by LLRs in Eq.~\ref{eq:time_release} \\
$K_1$&& A constant which describes the linear increase in the time up until which CO is released by LLRs in Eq.~\ref{eq:time_release} \\
$i_{ck}$ & & the smallest impactors that can cause catastrophic destruction of planetesimals in the $k$-th bin \\
$i_{rk}$ & & the smallest impactors that can cause re-surfacing collisions for planetesimals in the $k$-th bin \\
$i_{lrk}$ & & the largest remnant of a catastrophic collision in the $k$-th bin \\
$I$ & & the maximum inclination of colliders \\

$m_k$ & kg & The total mass of planetesimals in the $k$-th bin \\
$m_{s,k}$ & kg & The total mass of refractories (solids) in each planetesimal in the $k$-th bin \\

$m_{v,k}$ & kg & The total mass of volatiles in planetesimals in the $k$-th bin \\

$\dot{m_{\rm s,k}}$ &kgs$^{-1}$ & The mass loss rate of solids in the $k$-th bin\\

$\dot{m_{\rm s,k}}^{+c}$ &kgs$^{-1}$ & The mass loss rate of solids due to catastrophic collisions in the $k$-th bin\\

$\dot{m_{\rm s,k}}^{+c}$ &kgs$^{-1}$ & The rate at which the mass in solids in the $k$th bin changes due to catastrophic collisions \\

$\dot{m_{\rm v,k}}^{+c}$ &kgs$^{-1}$ & The rate at which the mass in volatiles in the $k$th bin changes due to catastrophic collisions \\
$\dot{m_{\rm s,k}}^{-r}$ &kgs$^{-1}$ & The mass loss rate of solids due to resurfacing collisions in the $k$-th bin\\

$\dot{m_{\rm v,k}}^{-r}$ &kgs$^{-1}$ & The mass loss rate of volatiles due to resurfacing collisions in the $k$-th bin\\

$\dot{m_{\rm v,k}}$ &kgs$^{-1}$ & The mass loss rate of volatiles in the $k$-th bin\\
$\dot{m}_g$ & & The total gas production rate\\

$\dot{m_{\rm g,k}}^{+c}$ &kgs$^{-1}$ & The gas production rate due to catastrophic collisions  in the $k$-th bin\\

$\dot{m_{\rm g,k}}^{-r}$ &kgs$^{-1}$ & The gas production rate due to resurfacing collisions  in the $k$-th bin\\

$\dot{m_{k, a}}$ & kgs$^{-1}$& The rate at which gas is relased due to the decay of LLRs \\ 
$m_{{\rm s, tot}}(0)$ & kg & The total mass in solids in the entire planetesimal belt at t=0 \\
$M_k$ & kg & The mass of each planetesimal in the $k$-th bin\\

$M_{ck}$ & kg & The mass of the smallest impactor that can cause a castrophic collision of planetesimals in the $k$-th bin \\
$M_{\rm min}$ && The minimum mass bin in the collisional cascade, equivalent to diameter, $D_{\rm min}$\\
$M_{\rm max}$ && The maximum mass bin in the collisional cascade, equivalent to diameter, $D_{\rm max}$\\

$M_{s,k}$ & kg & The mass of refractories (solids) in each planetesimal in the $k$-th bin \\

$M_{v,k}$ & kg & The mass of volatiles in each planetesimal in the $k$-th bin \\

$n_k$ &  & The number density of colliders in the $k$-th bin \\

$P_{ik}$ & & The intrinsic collision probability \\
$Q_D^*$ & Jkg$^{-1}$ & Dispersal threshold \\
$Q_S^*$ & Jkg$^{-1}$ &Shattering threshold \\

$Q_a$ & Jkg$^{-1}$ & Constant in the dispersal threshold Eq.~\ref{eq:qdstar}\\

$Q_b$ & Jkg$^{-1}$ & Constant in the dispersal threshold Eq.~\ref{eq:qdstar}\\

$\rho_k$ & kgm$^{-3}$ & Average density of planetesimals  \\

$\rho_{s,k}$ & kgm$^{-3}$ & Density of solid (refractory) component \\

$\rho_{v,k}$ & kgm$^{-3}$ & Density of icy component \\
$R_k^c$ & & The rate of catastrophic collisions in the $k$-th bin \\
$R_k^r$ & & The rate of re-surfacing collisions in the $k$-th bin \\

$r$ & au& The radius of the planetesimal belt\\
$dr$ & au & The width of the planetesimal belt \\ 
$t_{\rm release}$ & Myr& The time during which comets release CO from the decay of LLRs\\

$v_{rel}$ & ms$^{-1}$ & The average relative velocity of particle collisions \\
$V$ & & The volume through which planetesimals move and collide \\
$Z_k(t)$ & & A function which equals 1 for times, $t <t_{\rm release}^k$ and zero otherwise \\

\hline

\end{tabular}

\label{tab:variables}
\end{table*}

\begin{table}
\begin{tabular}{ccc}

\hline
Name & Age (Myr) & Mass in CO ($M_\oplus$) \\ \hline
  $\beta$~Pic &23&$2.8\times10^{-5}$ \\
49~Ceti     &40&$1.4 \times 10^{-4}$\\
HD~21997    &45&$6 \times 10^{-2}$\\
HD~95086    &15&$4.3 \times 10^{-6}$\\
HD~121617   &16&$1.8 \times 10^{-2}$\\
HD~131488   &16&$8.9 \times 10^{-2}$\\
HD~131835   &16&$4 \times 10^{-2}$\\
HD~138813   &10&$7.4 \times 10^{-4}$\\
   Fomalhaut&440&$6.5 \times 10^{-8}$\\
        TWA7&10&$8 \times 10^{-1}$\\
    HD 36546&10&$9 \times 10^{-2}$\\
  HD 129590&16&$1 \times 10^{-4}$\\
          HD~146897   &10&$2.1 \times 10^{-4}$\\
          HD~181327&23&$2.1 \times 10^{-6}$\\ \hline
\end{tabular}
\caption{ The mass of CO detected in main-sequence debris systems from Table A1 of \citet{Marino2020}, and additionally \citet{Matra2019TWA7}, \citet{Kral2019}, \citet{Rebollido2022}. }
\label{tab:CO}
\end{table}

\bibliographystyle{mn}

\bibliography{ref}


\bsp	
\label{lastpage}
\end{document}